\documentclass[12pt]{article}

\usepackage{graphicx}
\usepackage{cite}
\usepackage{amsmath, amssymb, amsfonts}
\usepackage{color}
\usepackage{latexsym}
\usepackage[normalem]{ulem}
\usepackage[labelfont=bf,margin=1cm]{caption}

\def\be{\begin{equation}}
\def\ee{\end{equation}}
\def\ba{\begin{eqnarray}}
\def\ea{\end{eqnarray}}

\def\bdm{\begin{displaymath}}
\def\edm{\end{displaymath}}
\def\la{~\mbox{\raisebox{-.6ex}{$\stackrel{<}{\sim}$}}~}
\def\ga{~\mbox{\raisebox{-.6ex}{$\stackrel{>}{\sim}$}}~}
\def\bq{\begin{quote}}
\def\eq{\end{quote}}

 at 10truept

\newcommand{\p}{\partial}

\def\ZZ{\mathbb{Z}}

\newcommand{\eps}{\epsilon}

\newcommand{\bea}{\begin{eqnarray}}
\newcommand{\eea}{\end{eqnarray}}

\newcommand{\half}{\frac{1}{2}}
\newcommand{\bi}{\begin{itemize}}
\newcommand{\ei}{\end{itemize}}

\newcommand{\beq}{\begin{equation}}
\newcommand{\eeq}{\end{equation}}
\newcommand{\beqa}{\begin{eqnarray}}
\newcommand{\eeqa}{\end{eqnarray}}

\def\la{~\mbox{\raisebox{-.6ex}{$\stackrel{<}{\sim}$}}~}
\def\ga{~\mbox{\raisebox{-.6ex}{$\stackrel{>}{\sim}$}}~}

\newcommand{\vev}[1]{\langle #1 \rangle}

\newcommand{\Z}{\mathbb{Z}}

\newcommand{\XX}{{\cal X}}
\newcommand{\MM}{{\cal M}}
\newcommand{\blam}{{\bar \lambda}}
\newcommand{\bg}{{\bar g}}
\newcommand{\cG}{{\cal G}}
\newcommand{\cB}{{\cal B}}

\def\ltap{\ \raise.3ex\hbox{$<$\kern-.75em\lower1ex\hbox{$\sim$}}\ }
\def\gtap{\ \raise.3ex\hbox{$>$\kern-.75em\lower1ex\hbox{$\sim$}}\ }
\def\gl{\ \raise.5ex\hbox{$>$}\kern-.8em\lower.5ex\hbox{$<$}\ }
\def\roughly#1{\raise.3ex\hbox{$#1$\kern-.75em\lower1ex\hbox{$\sim$}}}

\begin{document}

\thispagestyle{empty}
\begin{flushright}
June 2020 \\
BRX-TH 6665\\
\end{flushright}
\vspace*{.5cm}
\begin{center}

{\Large \bf On Hybrid Monodromy Inflation}\\
\vskip.3cm 
{\Large \it --- Hic Sunt Dracones --- }\\

\vspace*{.75cm} Nemanja Kaloper$^{a, }$\footnote{\tt
kaloper@physics.ucdavis.edu}, Morgane K\"onig$^{a, }$\footnote{\tt
mkonig@ucdavis.edu}, Albion Lawrence$^{b, }$\footnote{\tt
albion@brandeis.edu} and James H.C. Scargill$^{a, }$\footnote{\tt jhcscargill@gmail.com}\\
\vspace{.3cm}
{\em $^a$Department of Physics, University of
California, Davis, CA 95616, USA}\\
\vspace{.3cm} {\em $^b$Martin Fisher School of Physics, Brandeis University, Waltham, MA 02453, USA}\\

\vspace{1cm} ABSTRACT 
\end{center}

We revisit two-field hybrid inflation as an effective field theory for low-scale inflation with sub-Planckian scalar field ranges. We focus on a prototype model by Stewart because it allows for a red spectral tilt, which still fits the current data. We describe the constraints on this model imposed by current CMB measurements. We then explore the stability of this model to quantum corrections. We find that for relevant, marginal, and at least a finite set of irrelevant operators, some additional mechanism is required to render the model stable to corrections from both quantum field theory and quantum gravity. We outline a possible mechanism by realizing the scalars as compact axions dual to massive $4$-form field strengths, and outline how natural hybrid inflation may be supported by strong dynamics in the dual theory.

\vfill \setcounter{page}{0} \setcounter{footnote}{0}
\newpage

\section{Introduction}

Inflation is the leading candidate for explaining why the universe is so old, flat and smooth. Explicit models rely on a local
quantum field theory (QFT) of the agent behaving like a transient but long lived vacuum energy,  which forced the universe to rapidly expand early on. In a local Poincare-invariant QFT, this requires scalar fields with a very flat potential. An immediate question is whether such models are natural or fine-tuned. Leaving aside the conundrum of initial conditions, and merely focusing on the naturalness of the parameters of the local QFT, the inflationary epoch is obviously an arena where the ideology of naturalness may be tested observationally. 

The simplest QFT models of inflation involve a single scalar field which needs both a very flat potential and super-Planckian field variations \cite{Linde:1983gd,Linde:1987yb}. These large-field models typically produce observable primordial gravitational waves. Thus they are increasingly in tension with observations although they are not yet ruled out. However the super-Planckian field ranges also present a theoretical challenge, a large number of Planck-suppressed operators must be fine-tuned absent a mechanism for suppressing them \cite{Freese:1990rb,Adams:1992bn,Kim:2004rp,Silverstein:2008sg,Kaloper:2008qs,Kaloper:2008fb,McAllister:2008hb,Berg:2009tg,Dong:2010in,Kaloper:2011jz,DAmico:2012khf,Kaloper:2014zba,McAllister:2014mpa,Kaloper:2016fbr,DAmico:2017cda,Nomura:2017ehb,DAmico:2018mnx}. Since data is starting to take a bite out of these models, one may be tempted to avoid the issue of Planck-suppressed operators by studying small-field models, or by studying even more exotic alternatives where naturalness of the QFT of the inflaton is altogether abandoned. These latter alternatives are also not safe from the predations of quantum gravity. Therefore we will take a very conservative attitude, ignore exotic inflationary models altogether, and only consider inflationary small-field QFTs which might be natural. As we will find, such small-field models in fact require not so small fields after all, and are still threatened by Planck-suppressed irrelevant operators\footnote{This has been noted in similar contexts before, such as in \cite{Baumann:2010ys}.}. Some mechanism is still required to ensure that the quantum gravity corrections are under control.

In this paper we will revisit two-field hybrid inflation models \cite{Linde:1993cn,Copeland:1994vg,Binetruy:1996xj}, and explore their stability to both quantum field theoretic and quantum gravity corrections. The `classic' model of hybrid inflation \cite{Linde:1993cn,Copeland:1994vg,Binetruy:1996xj} predicts a blue spectrum of CMB perturbations that is now ruled out by observations \cite{Akrami:2018odb}. However, this does not shut the door  on hybrid inflation. For example, a variant due to Stewart \cite{Stewart:1994pt} (see also \cite{Lazarides:1995vr}) does produce a red spectrum (as we will review), and for some range of parameters can be made compatible with observations. We will find that additional quartic terms are induced by radiative corrections, which nevertheless can be kept naturally small. However -- unsurprisingly -- the quadratic mass terms suffer from the usual hierarchy problem afflicting fundamental scalars, and a mechanism is needed to keep them sufficiently small. Even with such a mechanism, we will find that the combination of data-based constraints and technical naturalness put the theory into a corner for which the range of the scalar fields are close in magnitude to the Planck scale. The main constraints are:
\begin{itemize}
\item We need to ensure that the inflaton rolls sufficiently slowly over a sub-Planckian range to generate $N$-efoldings of inflation. 
\item The two fields in the hybrid model can be roughly classified as ``inflaton" and ``waterfall" fields. The former rolls during inflation leading to a spectral index.  The latter is locked by the large value of the inflaton during inflation, and provides the bulk of the vacuum energy, and condenses at the end of inflation, ending vacuum dominance and initiating reheating. We ask that the field displacement of the ``waterfall" field is sub-Planckian, while being high enough to give the right scale of density perturbations when the couplings remain weak enough for radiative stability.  
\end{itemize}

The upshot is that two-field hybrid models that fit the data, are technically natural, and protected from quantum gravity corrections are very nontrivial to realize. In addition to requiring hierarchies in masses and renormalizable couplings, the nearly-Planckian field ranges mean that there are a large but finite number of Planck-suppressed irrelevant operators which may spoil slow-roll inflation unless there is a mechanism to make their dimensionless coefficients sufficiently small\footnote{In principle, it is possible that the individually dangerous operators may sum up to tamer contributions to the effective potential, if there are approximate shift symmetries in some regimes of phase space.}. Furthermore, satisfying the demands placed on inflation models with sub-Planckian field ranges puts additional pressure   on the EFT which one uses to describe the dynamics, because the inflaton -- which must be light during inflation -- quickly becomes very heavy after inflation ends. This is merely a matter of continuity: if the field rolls on a flat and short plateau to give $\ga 60$ efoldings of inflation, that plateau must be very flat, and the post-plateau minimum very narrow, and hence with a large curvature. This pushes reheating out of the EFT used to describe inflation: the UV embedding must be understood.

After exploring these issues, we will argue that an embedding of the Stewart model exists which can protect the required hierarchies in masses and couplings from both QFT and quantum gravity corrections. To do this we will use a two-field version of the axion monodromy effective field theories developed in \cite{Kaloper:2008qs,Kaloper:2008fb,Kaloper:2011jz,Kaloper:2014zba,Kaloper:2016fbr,DAmico:2017cda,DAmico:2018mnx}. In these theories the two fields that drive inflation and reheating are considered as axion-like pseudoscalars, dual to longitudinal modes of two massive $4$-form field strengths. The scalar mass maps to the gauge field mass \cite{Kaloper:2016fbr}; as with masses for Abelian vector fields, the $4$-form mass is stable to quantum corrections. Additional small hierarchies needed to support inflationary dynamics may then arise from dimensionless couplings in the strong coupling regime. For models which match CMB data one must either tune the value of the mass or find an additional mechanism, beyond EFT, to explain its smallness
from first principles. Either way, the small mass is at least technically natural\footnote{Lower bounds on the mass of $1$-form gauge fields have been conjectured in \cite{Reece:2018zvv}: we believe inflation is well away from such bounds.}. We will also 
see that the requirements of naturalness and bounds from data push the dual $4$-form theory into the strong coupling regime. 
However since the theory has only longitudinal propagating modes, 
in the inflationary regime the theory looks weakly coupled because each explicit power of the inflaton $\phi$ comes with a factor $\propto \mu/{\cal M}$, where 
${\cal M}$ is the cutoff and $\mu$ the inflaton mass, thus providing additional suppression factors that make the theory appear weakly coupled in the axial gauge, and also tame quantum gravity corrections to the irrelevant operators of the theory. 

The current data put additional pressure on this EFT, suggesting that a full theory of hybrid inflation really requires two different EFTs to consistently describe it,
one for the slow roll regime and another for reheating, when the inflaton becomes very heavy (and presumably other degrees of freedom
become very light). Providing the details of the complete first-principles construction of such a model 
is beyond the scope of this work. Here our approach is bottom-up, 
reverse-engineering a low energy theory in order to glimpse how it can be derived from a UV-complete model. Hence we cannot provide a detailed contents of the full spectrum of the theory, which is necessary to see how various fields transition to below and above the cutoff. Nevertheless, 
our analysis indicates that such constructions exist, and serves as a guide for how to search for more complete models which realize such dynamics. 

We close with a summary of what is new in this paper:
\begin{itemize}
\item First, a detailed consideration of matching the Stewart model to the most up-to-date data has not been done before; nor has the stability of this model to either QFT or quantum gravity corrections  been explored. We do both in \S3. While the model fits data well, we will find that as it stands, it in fact is unnatural: the model suffers from the mass hierarchy problem, since the dimensional scales in the theory are very sensitive to the UV completion. This requires at least a small ${\cal O}(0.1)$ hierarchy between mass scales. A protection mechanism is necessary to stabilize this against quantum corrections and render the theory at least technically natural. We will also see that the 
theory has many irrelevant operators which are individually large, and a dynamical explanation of why they 
do not spoil inflation is needed. 

\item Furthermore, we will also see that the theory is additionally strained by the fact that the post-inflationary mass of the inflaton is near the cutoff of the infationary EFT, meaning that to reliably describe exit from inflation and reheating, we really should use a different EFT in that regime.

\item Finally, our proposal for realizing the model as the theory of coupled axions dual to $4$-form field strengths is new. This provides a UV completion of the hybrid inflation model which addresses the hierarchy problem and the subsequent fine tunings, rendering it both natural and protected from quantum gravity corrections. To do this we take an approach that is distinct from standard axionic model-building. Typically axions have not been considered for hybrid models because their nonderivative couplings are constrained by nonrenormalization theorems, and taken to be simple trigonometric functions. This assumes that the axion potentials are generated by instantons in some sort of dilute gas approximation. However, it has been known for some time, if underappreciated, that this approximation often fails and axion potentials can be more complex, looking more like perturbative polynomial interactions, if we assume that the axion potential is multivalued \cite{Witten:1978bc,Witten:1980sp,Witten:1998uka}.  The central lesson of \cite{Kaloper:2008qs,Kaloper:2008fb,Kaloper:2011jz,Kaloper:2014zba,Kaloper:2016fbr,DAmico:2017cda,DAmico:2018mnx}\ is that the resulting effective field theories are still constrained by gauge symmetries, and `dimensional transmutation' between dual pictures allows for small couplings that are protected from QFT loops and quantum gravity corrections. In \S3\ we develop and deepen this story by expanding it to multifield models and explore its strong coupling regime where inflationary dynamics may be realized. In addition to the application at hand, our hope is that we are expanding the toolkit for QFT models more generally.

\end{itemize}

\section{Hybrid inflation with a red spectrum}

\subsection{The model}

The pioneering model of hybrid inflation \cite{Linde:1993cn} has the form:
\begin{equation}
V(\phi, \sigma) = \frac{\lambda}{4} \left( \sigma^{2} - \tilde M^{2}\right)^{2} +\half \mu^{2}\phi^{2} + \frac14 g^2 \sigma^{2}\phi^{2} \, .
\end{equation}
Here $\phi$ is the inflaton. If $\phi^2 > \lambda \tilde M^2/g$, then $\sigma = 0$ is a minimum of the potential for fixed $\phi$, leading to the effective potential
\be
	V_{eff} = \tilde M^4 + \half \mu^2 \phi^2 \, .
\ee
During inflation, if the second term dominates, the theory is effectively the standard chaotic inflation, which requires a super-Planckian range for $\phi$ and which is ruled out by existing bounds on gravitational waves.  If the $M^4$ term dominates the potential, then
\be
	\left(\frac{\delta\rho}{\rho}\right)^2 =  \frac{V^3}{24\pi^2 M_{pl}^6 (V')^2} \sim \frac{1}{\phi^2} \, ,
\ee
grows larger as $\phi$ decreases, leading to a blue spectral index, which is ruled out by CMB measurements  \cite{Akrami:2018odb}. More precisely the spectral index is
\be
	n_s = 1 + \frac{2}{N} > 1 \, , 
\ee
where $N$ is the number of efoldings.

An alternate model by Stewart \cite{Stewart:1994pt,Lazarides:1995vr}, defined by
\be
	V(\phi,\sigma) = \frac{m^2}{2}\left(\sigma - \sqrt{2}M\right)^{2} + \frac{g^{2}}{4}\phi^{2}\sigma^{2}\, , \label{eq:stewartmodel}
\ee
which does not have degenerate post-inflationary minima, and so also no dangers of any stable defect production after inflation, does have a red spectrum:
\be
	n_s \sim 1 - \frac{3}{2N} \, .
\ee
As we will show in \S2.2, this model can be made compatible with current CMB data, while maintaining sub-Planckian field ranges that are one of the strong motivations for hybrid inflation.

As far as we know, a full discussion of the quantum stability of these models has never been carried out\footnote{We are aware of a somewhat incomplete attempt  in \cite{Lyth:1999sp}.}. This is one of the main goals of this paper.  We thus write a modification of Eq. (\ref{eq:stewartmodel}) containing all of the relevant and marginal operators that could be generated by quantum corrections:
\be 
	 V(\phi,\sigma) = \frac{m^2}{2}\left(\sigma - \sqrt{2}M\right)^{2} +\frac{\lambda}{4}\left(\sigma^{2} - \tilde{M}^{2} \right)^{2} +\frac{\mu^2}{2}\phi^{2}+\frac{g^{2}}{4}\phi^{2}\sigma^{2} + \frac{\lambda'}{4} \phi^4 \, .
	 \label{potpot}
\ee
Since the first term breaks the $\Z_2$ symmetry $\sigma \to - \sigma$ one may be tempted to include terms proportional to $\sigma \phi^2$ or $\sigma^3$. However, if we compute quantum corrections to this potential following \cite{Coleman:1973jx}, linear terms in $\sigma$ can be absorbed into the source term used to explore the full theory space, and thus these terms are never generated by loop corrections. In the effective field theory (EFT) language, the reason is that the parity symmetry is softly broken, only by a relevant operator, and thus is invisible to the UV effects in the loops.

In the EFT context, we would naturally expect that all of the dimensionful parameters in this model should be below the cutoff. However, we will find in the next section that for a realization of this model via massive $4$-forms, this does not have to be the case for $M$, for which there is a see-saw formula involving the ratio of the cutoff to a fundamental mass scale. A hint that this might be the case comes from noting that if $\sigma$ is a pseudoscalar axion, as it (along with $\phi$) is a dual of a massive $4$-form field strength in the embedding we described below, $M$ is a spurion for the breaking of CP, and so it might naturally be related to the initial expectation value of one of the longitudinal modes of massive dual forms; we will see that this is so in \S3. There is no a priori reason that the expectation value of an axion field should be below the cutoff \cite{Kaloper:2015jcz}. In principle, since it is controlled by the initial flux of a $4$-form field strength it may even be $\sim M_{pl}$. 

The first and fifth terms are absent in \cite{Linde:1993cn}; while Ref. \cite{Stewart:1994pt}\ ignores the second, third and fifth terms.  To realize the latter model with its red spectrum of fluctuations, we must ensure that the terms controlled by $\mu, \lambda,\lambda'$ are subdominant, so that the Stewart model is a good approximation.  We will argue in \S2.3\ that the suppression of $\lambda,\lambda'$ can be made technically natural. Finding a phenomenologically acceptable value of $m$ and suppressing $\mu$ enough that the $\phi^2$ term is subdominant requires more work, as scalar masses are afflicted by the hierarchy problem.  Rendering these masses even technically natural requires a mechanism, such as axion monodromy, whose application to hybrid inflation we will describe in \S3.

For now and in \S2.2, we will take (\ref{potpot})\ as given and explore the predictions of the model in the limit well described by the Stewart model. As in \cite{Linde:1993cn,Stewart:1994pt}, at sufficiently large $\phi$, the $g^2 \phi^2\sigma^2$ term renders $\sigma$ massive and we can integrate it out.  At the classical level we simply solve the equation 
\be
	\frac{\partial V}{\partial \sigma} = m^{2}\left( \sigma - \sqrt{2}M \right)+\lambda \sigma^{3} -\lambda\tilde{M}^{2}\sigma +\frac{g^{2}}{2}\phi^{2}\sigma = 0 \, ,
	\label{minpot}
\ee
for $\sigma$ at large $\phi$.  We demand that the term $\lambda\sigma^3$ be subdominant. Ignoring this term, we find that:
\be
	\sigma_{min} = \frac{\sqrt{2}m^{2}M}{m^{2}-\lambda\tilde{M}^{2} +\frac{g^2}{2}\phi^{2}} \, .
	 \label{sigmin}
\ee

It is instructive to neglect the term $\propto \lambda$, and in this limit plug equation (\ref{sigmin}) back into (\ref{potpot}). So, using
$\sigma_{min} = {\sqrt{2}m^{2}M}/({m^{2}+\frac{g^2}{2}\phi^{2}}) + {\cal O}(\lambda)$, this yields
\be 
	 V_{eff}(\phi) \simeq {m^2}M^2  \frac{\frac{g^2\phi^2}{2}}{{m^{2}+\frac{g^2}{2}\phi^{2}}} + \frac{\mu^2}{2}\phi^{2} + \frac{\lambda'}{4} \phi^4 + {\cal O}({\lambda})\, .
	 \label{potpotinfl}
\ee
If $ \phi^2 \gg 2 m^2/g^2$, then $\sigma \simeq \frac{2\sqrt{2} m^{2}M}{g^{2}\phi^{2}}$, and in this limit the potential (\ref{potpotinfl}) 
becomes 
\be
	\label{eq:inflation}
	V_{inflation} \simeq m^{2}M^{2}\left(1-\frac{2m^{2}}{g^{2}\phi^{2}}\right) + \frac{\mu^{2}}{2}\phi^{2} + \frac{\lambda'}{4} \phi^4 + {\cal O}({\lambda}) \, ,
\end{equation}
after expanding the fraction in the first term. Clearly, as $\phi$ grows the first term becomes flatter. This is the inflationary plateau. The terms $\propto \mu, \lambda'$ limit it, since they make 
the potential convex again as they take over. Essentially, the regime with positive spectral slope is therefore between the two inflection points of the potential
(\ref{potpotinfl}), when $\phi$ is in the regime
\be
\sqrt{\frac23} \frac{m}{g} \le \phi \le (12)^{1/4} \sqrt{\frac{M}{g \mu}} m  \, .
\label{inflreg}
\ee
The hybrid inflation regime, with sub-Planckian ranges for $\phi$ and a red spectrum, will require that the terms controlled by $\mu,\lambda'$ also be sufficiently subdominant, during the first 10 efoldings of the visible epoch of inflation (epochs which leave imprints on the CMB).  The precise details depend on how small $\mu$ is. If $\mu$ is larger than the critical value 
$\mu_* = \sqrt{12} \frac{Mm^2}{g M_{pl}^2}$, the upper bound in (\ref{inflreg}) is 
sub-Planckian, and so the period of inflation with a red spectrum is generically shorter with all other parameters being fixed. 
 If $\mu$ is smaller than $\mu_*$, the upper bound is super-Planckian. At any rate we will insist that the final $50$ efoldings must occur over a sub-Planckian range of the inflaton, while the inflaton lies below the upper bound in (\ref{inflreg}).  In other words, if we define the maximum value of $\phi$ at $50$ efolds before the end of inflation as $\alpha m_{pl}$, we require $\alpha <1$. The upper bound in $\phi$ in (\ref{inflreg}) then implies $\sqrt{12} Mm^2 \ga \alpha^2 g M_{pl}^2 \mu$.  
In this way, whichever option, 
any trans-Planckian field excursions are observationally irrelevant, offering assurance that our model of nature need not be too sensitive to the UV. We will however see that this puts stress on naturalness. 

We will further require that $\mu/m <1$, so that we can consistently integrate out the $\sigma$ field during inflation,
and ignore its fluctuations. This also ensures that the $\sigma$ fluctuations during inflation are suppressed; as a consequence we can ignore isocurvature perturbations, which are  anyway strongly constrained by
the data. We will however find that when other bounds are met, $\mu/m <1$ is automatically satisfied. 

The limitations of an EFT are set by the cutoff ${\cal M}$.  While the effective potential (\ref{potpot}) involves only renormalizable operators, and so it gives no indication of any intrinsic UV cutoff, we can place a lower bound on the cutoff of the two-field model by noting that ${\cal M}$ should be
of the order of the effective mass of $\sigma$ during inflation. This mass will provide a natural UV cutoff for the effective theory, Eqs. (\ref{potpotinfl}) and (\ref{eq:inflation}) for $\phi$, once we integrate out $\sigma$. This potential is obtained for large $\phi$ by integrating the field $\sigma$ when it is effectively near zero. Taking the second derivative of (\ref{minpot}) gives the effective mass of $\sigma$ in that regime, 
\be
m^2_\sigma \simeq m^2 +\frac{ g^2 \phi^2}{2} + {\cal O}(\lambda) \, .
\label{sigmass}
\ee
To get an idea about the limit on the inflaton $\phi$ range in our EFT we conservatively use the range $\sim M$ of $\sigma$ as a guideline\footnote{This assumes a lack of hierarchy between $\sigma$, $\phi$; in practice we expect $\phi$ to have a somewhat larger range, as we are working to flatten its potential.}. Imposing $M < M_{pl}$, we thus take the cutoff ${\cal M} = \sqrt{4\pi} {\cal M}_*$ to be 
${\cal M} \sim g M$. Here ${\cal M}_*$ is the strong coupling scale, used to normalize the EFT 
operator expansion in the Na\"ive Dimensional Analysis (NDA) framework \cite{Manohar:1983md,Gavela:2016bzc}, a framework which was deployed to organize the EFT of large field inflation in \cite{Kaloper:2016fbr,DAmico:2017cda}. To get a sharper estimate of the cutoff, we  can also produce a lower bound by recalling that in NDA, the overall dimensional scale of the potential terms in EFT is ${\cal M}_*^4$ at strong coupling. This sets an upper bound $V \la {\cal M}_*^4$ on the potential energy during inflation \cite{Kaloper:2016fbr,DAmico:2017cda}. From (\ref{eq:inflation}), we then find ${\cal M}_* \ga \sqrt{mM}$. Thus the cutoff lies in the range
\be
\sqrt{4 \pi mM} \la {\cal M} \sim g M \, .
\label{scalcutoff}
\ee
For values of these parameters which satisfy constraints we develop below, we find $gM > \sqrt{mM}$ so we take $gM$ as a conservative estimate of ${\cal M}$.

Note, however, that as the inflaton $\phi$ rolls towards the end of inflation,
its effective (tachyonic) mass on the plateau, $|m_\phi^2| \sim 24 \frac{m^2}{{\cal M}_*^2} m^2 \ll m^2$ changes to 
\be
m_{\phi}^2 \sim g^2 M^2  \sim {\cal M}^2 \, .
\label{phimass}
\ee
Thus the EFT of inflation and the stage right after inflation breaks down 
as the theory proceeds to reheating at the true vacuum $\sigma = \sqrt{2} M$. 
As serious as this is, by itself it is {\it not} 
a fundamental flaw of the theory. It indicates that as inflation nears the end, the theory is
undergoing a phase transition where the energy density stored in the very flat potential tends to dissipate quickly, and nonperturbatively,
at very short distances. Some new fields, including $\sigma$ -- which was very heavy during inflation -- become light and need to be integrated in. Others, like the inflaton $\phi$, become 
heavy. Further, since the $\phi$ mass is tachyonic, as its magnitude becomes larger the tachyonic instability becomes
faster near the end of inflation and operates at sub-horizon scales. So the 
important lesson from this is a {\it warning}: we should not expect to have a single EFT of hybrid inflation,
but instead count on having the late stages of the exit and reheating as separate descriptions from slow roll inflation.

\subsection{Matching to data}

We next want to demonstrate that our model (\ref{potpotinfl}) has the capacity to match existing CMB data, focusing on:
\begin{itemize}
\item The number $N$ of efoldings of inflation, constrained (up to details of reheating, which we do not model here) by bounds on spatial curvature to be at least on the order of $50-60$. For the sake of convenience we take $N = 50$.
\item The power in scalar density (or CMB temperature) fluctuations at an appropriate pivot point, set by experiments to be of order $\frac{\delta\rho}{\rho} \sim 5\times 10^{-5}$.
\item The tensor-scalar ratio $r$, with a current upper bound at $r \leq 0.056$ \cite{Akrami:2018odb}. We will in fact demand here that the inflaton $\phi$ has a sub-Planckian field range over the last $N$ efoldings of inflation; this ensures that $r$ is well below the above bound \cite{Lyth:1999sp,Efstathiou:2005tq}. 
\item The scalar spectral index $\alpha_S$, between $0.95$ and $0.985$, depending on $r$.
\end{itemize}

We will do this here under the assumption that  the first two terms in (\ref{eq:inflation})\ dominate in our model. The above constraints can then be phrased in terms of bounds on $m,M,g$ and the maximum excursion $\alpha M_{pl} \equiv \phi_{max}$ of the scalar field during the visible epoch. We will find that the data supports a range for these parameters. The further question of whether there is a range for these parameters and for $\lambda, {\tilde M},\mu,\lambda'$ for which the remaining terms in (\ref{eq:inflation}) are subdominant and the whole theory is stable under quantum corrections is the central point of the rest of this paper.

We first impose the constraint that inflation last for a sufficient number of efoldings. Assuming that the magnitude of the potential energy $V$ is dominated by the first term in (\ref{eq:inflation}), the number $N$ of efoldingsß is given by:
\begin{equation}
N = - \frac{1}{M_{pl}^2}\int \frac{V}{V'}\mathrm{d}\phi \simeq \frac{1}{M_{pl}^{2}}\frac{g^2}{16m^{2}}\phi_{max}^{4} \, , \label{eq:efolds}
\end{equation}
Using $\phi_{max} = \alpha M_{pl}$, we find:
\be
	\frac{m}{g M_{pl}} \simeq \frac{\alpha^2}{4 \sqrt{N}} \, . \label{eq:efconstr}
\ee
Note that we are interested in $\alpha \la 1$; thus, $m \ll g M_{pl}$. Since we also wish to impose a sub-Planckian range of $\sigma$, and that range is set by $M$, this condition is at least compatible with $m < {\cal M} \sim g M$.

Our bound assumed that $V$ is dominated by the constant term $\sim m^2 M^2$ during the first $\sim 10$ efoldings of the visible epoch of inflation (the period which leads to observed CMB fluctuations). We need to check that this is self-consistent. $10$ efoldings corresponds to  to approximately $\frac{\p\phi_{max}}{\p N} \Delta N \sim \frac{2.5}{N} \phi_{max}$ which is a small variation, so we need simply check the dominance of the constant term  at $\phi \sim \phi_{max}$. Under these conditions, 
\be
\frac{m}{g\phi_{max}} = \frac{m}{\alpha g M_{pl}} \simeq \frac{\alpha}{4\sqrt{N}} \ll 1 \, , \label{eq:sigmamassconstr}
\ee
so our assumption of the dominance of the constant term in $V$ is self-consistent. Note that in light of the discussion in the previous section, this demonstrates that for parameters that yield the requisite number of efoldings, $m \ll {\cal M}$.  Also note that it is straightforward to show that the slow roll parameters $\eta \sim M_{pl}^2 V''/V$, $\eps \sim M_{pl}^2 (V'/V)^2$ are both small given (\ref{eq:efconstr}) together with $\alpha \la 1$.

We next check the scalar spectral index:
\begin{equation}
n_{S} = 1-3M_{pl}^{2}\left(\frac{V'}{V}\right)^{2} + 2 M_{pl}^{2}\frac{V''}{V} = 1-24M_{pl}^{2}\frac{m^{2}}{g^{2}\phi^{4}}\left(\frac{2m^{2}}{g^{2}\phi^{2}} +1\right) \simeq1- \frac{3}{2N} \, .
\end{equation}
In the last term in the RHS we are assuming that $\frac{2 m^2}{g^2\phi^2} \ll 1$: this is the same as the demand that $V$ in (\ref{eq:inflation}) is dominated by the leading constant term $m^2 M^2$. If we take $N$ between $50$ and $60$, $n_s$ varies between $0.97$ and $0.975$. This is within current bounds, particularly for $N \sim 50$. However future observations might be able to constrain this more strongly, and perhaps even falsify the model.  In any case, as we noted, in the rest of this paper we will take $N=50$ as the pivot point to match the model to the data. 

Thirdly, we impose a constraint on $m,M,g,\alpha$ from the power in scalar fluctuations:
\begin{equation}
\frac{\delta \rho}{\rho} = \frac{1}{2\pi}\frac{H^{2}}{\dot{\phi}} =\frac{1}{2\pi\sqrt{3}M_{pl}^{3}}\frac{V^{3/2}}{V'}=
\frac{g^{2} M \phi^{3}}{8\pi\sqrt{3} m M_{pl}^{3}} \, ,
\label{contrast}
\end{equation}
where we assume, following the above discussion, that the first term in (\ref{eq:inflation})\ dominates the magnitude of $V$.
Using Eqs. (\ref{eq:efconstr}) and (\ref{contrast}), and that $\delta \rho/\rho \approx 5 \times 10^{-5}$, we find two constraints 
on $m, M, g, \alpha$. We can express any two in terms of the other two. For convenience, we will use these 
equations to express $m$ and $M$ in terms of the two remaining parameters $g$ and $\alpha$, which
we will treat as independent parameters, at $N=50$. We find
\be
\label{eq:data_constraints}
	\frac{m}{ M_{pl}} \simeq \frac{\alpha^2 g}{28} \, , ~~~~~~~~~~
	\frac{M}{M_{pl}} \simeq  \frac{8 \cdot 10^{-5}}{\alpha g}  \, ,
	\ee
where we substituted the first equation into (\ref{contrast}) to obtain the second one. We note that the parameters
$g$ and $\alpha$ cannot be  completely arbitrary. In addition to $g<1$ and $\alpha \la 1$, their choice 
must be made to maintain $M/M_{Pl} \la 1$, and ensure that the approximation where we neglected
$\mu$ and $\lambda'$-dependent terms is self-consistent, even when we include quantum corrections.
We will see that those requirements yield non-trivial restrictions on $g$ and $\alpha$.

Finally, the tensor-scalar ratio $r$ is given by:
\begin{equation}
r = 6M_{pl}^{2}\left(\frac{V'}{V} \right)^{2} = 6M_{pl}^{2}\frac{16m^{4}}{g^{4}\phi^{6}}=\frac{3m}{2gM_{p}}\frac{1}{N^{3/2}} \, .
\end{equation}
Using (\ref{eq:efconstr}), we find
\be
	r = \frac{3 \alpha^2}{8 N^2} \simeq 1.5 \times 10^{-4} \alpha^2 \, ,
\ee
well below the current bounds, and likely unobservable, given $\alpha < 1$.
Thus we see that if, e.g, $r$ were observed, and if $n_S \sim 0.95$, our model would be challenged by data.

Using the above constraints, we wish to display more explicitly the region in the space of parameters $m, M, g, \alpha$ that satisfy them. First,
the scale of the potential, and therefore the lower limit (\ref{scalcutoff})\ on the cutoff ${\cal M}$ is
\be
	V^{1/4} \simeq \sqrt{m M} \simeq \sqrt{\alpha} \times 10^{-3} M_{pl} \la {\cal M}/\sqrt{4\pi} \, .
	\label{mscale}
\ee
As we noted above, we will impose the requirement that all the fields remain sub-Planckian, in addition to the dimensional parameters
that appear in the relevant operators of the theory.  The field range for $\sigma$ is $\sim \sqrt{2} M$ so we demand $M < M_{pl}$ as well as $\alpha < 1$.  This is what helps keep the tensor power low. Furthermore, in principle this might have helped with keeping quantum gravity effects under control and allowing inflation and reheating to take place in the same effective field theory. As have seen and will see, these conditions are sufficient for neither. We have already argued that reheating takes place through a phase transition during which energies and momenta of order the cutoff become activated. We will show below that the Stewart model will be at best barely sub-Planckian, so that a large number of irrelevant operators must have small coefficients, even if they appear as Planck-suppressed.
Note also that when it is helpful to gain a conceptual handle on the constraints of our theory, we will impose a field space ``democracy'' with $\sigma, \phi \la M$ up to ${\cal O}(1)$ factors. 

With these in mind, we find that the second equation in (\ref{eq:data_constraints}) combined with $M/M_{pl}<1$ yields a bound
\be
g \ga \frac{8 \cdot 10^{-5}}{\alpha } \, .
\label{gbound1}
\ee
It is worth noting that this bound makes $\mu/m < 1$ consistent. This is needed for $\phi$ to be a good candidate for the inflaton with the fluctuations of $\sigma$ suppressed during inflation. In particular if we impose $\mu < \mu_*$, so that the plateau lasts for at least a Planck scale in range before quantum gravity cuts it off, then combining $\mu_* = \sqrt{12} \, \frac{Mm^2}{g M_{pl}^2}$ with the first of Eqs. (\ref{eq:data_constraints}) and using inequality (\ref{gbound1}) gives $\mu_* \simeq 3.4 \times 10^{-7} \alpha M_{pl}$, and so, using (\ref{gbound1}), 
\be
\frac{\mu}{m} < 0.1 \, .
\label{mubound}
\ee

Consistency of our EFT also demands that $m \ll {\cal M} \sim gM$ (using Eq. (\ref{scalcutoff})). Combining  the two Eqs. (\ref{eq:data_constraints}) we find $m/M = 500 \, \alpha^3 g^2$,  and thus $m/gM = 500 \, \alpha^3 g$. 
We already noted however that $m/gM \sim m/{\cal M} \ll 1$,
and will recheck it in the next section that requiring radiative stability of the marginal operators in the theory
yields an independent bound $g \la 1.6 \times 10^{-3}$, which guarantees $m/gM < 1$ as well as $m \ll M$.

In the end there is a nontrivial subspace of the parameters $m, M, g, \alpha$ satisfying our bounds.
In Fig. 1 we have plotted a region in the space of $g, \alpha$, which as we noted we treat
as independent parameters. For completeness we have also included the bound $g \la 1.6 \times 10^{-3}$, which we will
derive in the next section. Suppressing corrections to make the theory radiatively stable favors weaker couplings, $g <1$. However, 
as seen from Eq. (\ref{gbound1}), $g$ cannot be arbitrarily small as long as $\alpha <1$. Decreasing $g$ mandates increasing $M$ and/or $\alpha$ to keep $\delta \rho/\rho$ fixed. Conversely, 
lowering $\alpha$ and/or $M/M_{pl}$ puts pressure on the perturbativity of $g$.  
The theory therefore does {\it not} really work at arbitrarily low scales, as is clear from Fig. 1. Nonetheless, there is still a nontrivial window here; in particular, for e.g. $\alpha, M/M_{pl} \sim 0.1$, we can have $g \sim 0.001$, and so on. 

\begin{figure}[th]
\label{diagram}
\centering
\includegraphics[height=8.5cm]{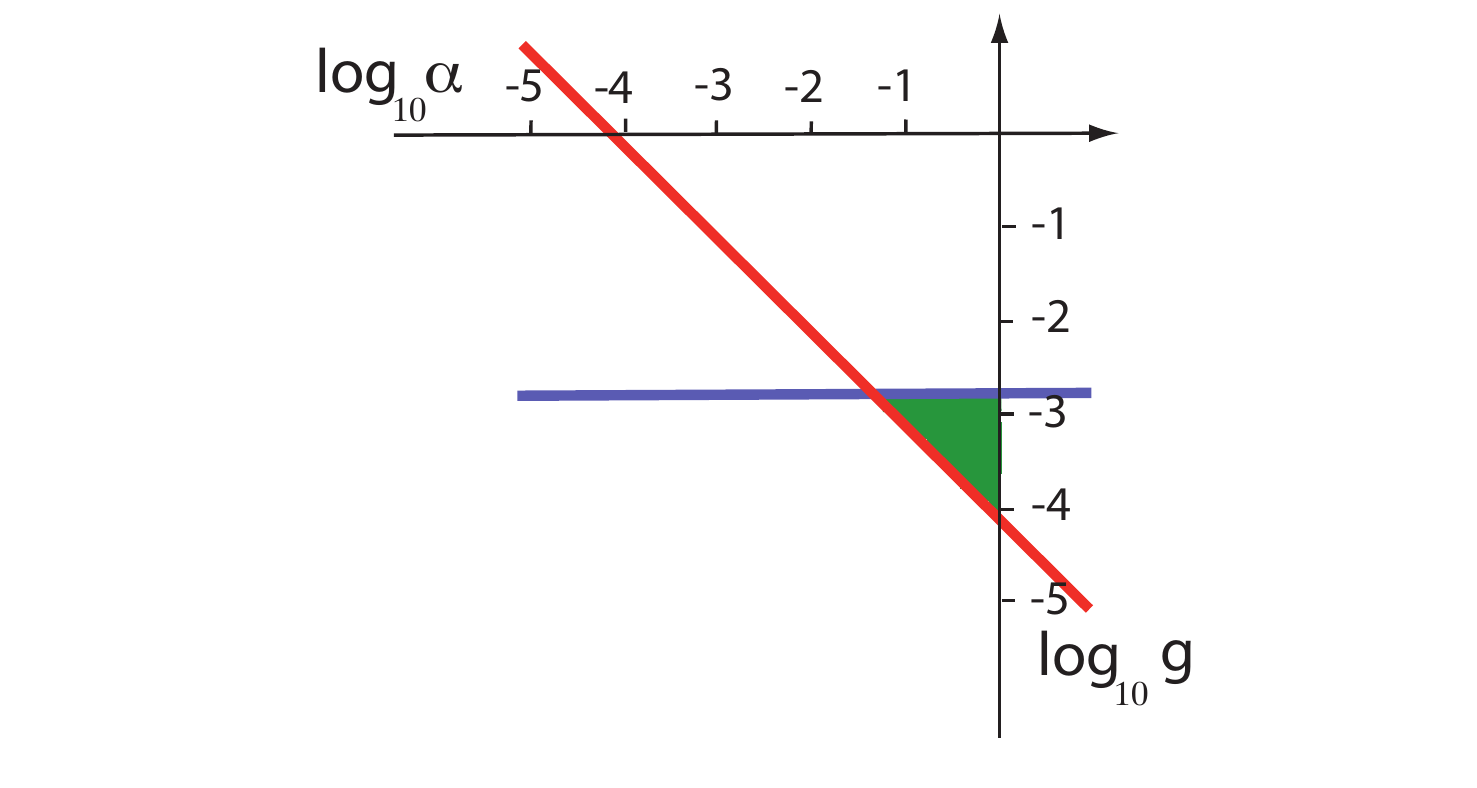}
\caption{\small Constraints on the inflation parameters $g,\alpha$ due to data and the naturalness of marginal operators. The inclined line corresponds to the bound of Eq. (\ref{gbound1}), which follows from $M/M_{pl}  \le 1$. The vertical coordinate 
axis is the bound $\alpha \le 1$. The horizontal bound comes from 
$g \le 1.6 \times 10^{-3}$ which follows from imposing naturalness of quartic operators (see \S 2.3.2). The green shaded region is the regime for which the Stewart model is consistent with all of these bounds. Note that this figure ignores the bounds from naturalness of the mass terms, which are problematic for the Stewart model (\ref{eq:stewartmodel}). The reason we are ignoring these bounds is because ultimately they may be addressed  by embedding (\ref{eq:stewartmodel}) in dual form monodromy, as we will outline in \S 3.}
\end{figure}

We forewarn the reader that in the plots in Fig. 1 we have ignored the ``elephant in the room" of the
Stewart model (\ref{eq:stewartmodel}): the power law divergent contributions to the relevant operators, specifically the mass terms. 
As we will argue, these require additional fine tunings, unless a mechanism is found which cancels them. Our goal in this paper is to indeed provide an example of such a mechanism, an embedding of (\ref{eq:stewartmodel}) into a theory of dual forms, so that $\phi,\sigma$ are pseudoscalars exhibiting axion monodromy. However even after the cancellation, the bounds discussed here remain. In particular we stress that the EFT of hybrid inflation can be \underline{at best} barely sub-Planckian in order to be both technically natural and consistent with data. The specter of quantum gravity still haunts hybrid inflation.

The reader may be disturbed that $M$ may well be the UV cutoff of many string compactifications. As we have discussed above and will expand on in more detail in \S3, this does not place the theory outside of the bounds of effective field theory below said cutoff.

\subsection{Quantum stability}

In this section we discuss the degree to which the Stewart model (\ref{eq:stewartmodel}) can be rendered technically natural.  The model will pick up corrections that include all of the terms in (\ref{potpot}). If we take the Stewart model as the tree-level action, the corrections will be generated through loops by the coupling $g^2$.  We will thus assume that the additional relevant and marginal couplings in (\ref{potpot}) that are absent from (\ref{eq:stewartmodel}) will be of the order of these loop corrections. The field theory corrections to irrelevant terms are also dangerous. 
For these terms, we will note that quantum gravity is expected to generate Planck-suppressed operators with order ${\cal O}(1)$ coefficients in a theory of quantum gravity, and we will find that these dominate the corrections generated by QFT loops.  We will then show that despite the sub-Planckian field ranges of the Stewart model, a large but finite number of irrelevant operators must have small coefficients in order for the Stewart model to be a good description of the dynamics, so that an additional mechanism or fine tuning is needed to suppress irrelevant operators. In \S3, we will show that axion monodromy can provide it, as it does for large field inflation  \cite{Kaloper:2011jz}.

The relevant terms in the Stewart model comprise the linear term $\propto m^2 M \sigma$, which does not get corrected, as we have discussed, and the scalar masses $m,\mu$.  As with the Standard Model, the masses suffer from a potential hierarchy problem.  We can realistically expect field-theoretic corrections to give $\delta m^2,\delta \mu^2 \sim \frac{g^2}{16\pi^2} {\cal M}^2$ where ${\cal M}$ is the cutoff, under the generous assumption that $\phi, \sigma$ couple with strength $g$ to ultraviolet degrees of freedom. This encodes the well-known ultraviolet sensitivity of the masses. There is also a weaker version of the hierarchy problem, which is that the loops of the heavy field $\sigma$ can correct the light mass of $\phi$ by large terms. 

In general, we can tame these corrections to a certain degree if we assume that a softly broken shift symmetry holds up to the fundamental scale in the limit $g\to 0$. This of course is technical naturalness. Current lore is that {\it nonperturbative}\  quantum gravity effects spoil such symmetries.\footnote{Perturbative quantum gravity effects preserve the shift symmetry of scalars: see \cite{Linde:1987yb,Kaloper:2011jz}.} An open question is whether these effects include relevant operators. As case studies of related Euclidean wormhole effects on the Peccei-Quinn symmetry of axion potentials
show, the complete resolution of these questions remains open \cite{Kallosh:1995hi,Kamionkowski:1992mf} (see also \cite{Hebecker:2016dsw,Daus:2020vtf}).  All we will do for this paper is merely parameterize the problem, leaving the issue of the existence of softly broken 
symmetries conceptually unresolved. We will instead argue in \S3 that gauge symmetries of the dual formulation in terms of massive
$4$-form field strengths may accomplish the same goal. 

\subsubsection{Scalar masses}

Mass terms for $\phi,\sigma$ at the cutoff ${\cal M}$ would clearly invalidate our EFT, pushing all dynamics to the cutoff. Furthermore, if we estimate ${\cal M} \sim g M$, we have shown that constraints from data imply that $m/{\cal M} \sim 500 \alpha^3 g$.  If we impose the constraints on $g \lesssim 10^{-3}$ from technical naturalness, and demand that inflation occur over a sub-Planckian field range, then some mechanism must keep $m$ well below the cutoff. Furthermore, we must also impose $\mu \ll m$.  If we require that the second derivative of $\mu^2 \phi^2$  be subdominant to the second derivative of the second term in (\ref{eq:inflation}), then, if $m,\mu \sim {\cal M}$, we find that this condition combined with (\ref{eq:efconstr}) implies that 
\be
	M \gg \left(\frac{\mu}{m}\right) \times  \sqrt{\frac{4 N}{3}} M_{pl} \sim \left(\frac{\mu}{m}\right) \times 8 M_{pl}
	\label{eq:masshier}
\ee
so that a sub-Planckian field range for $\sigma$ requires $\mu <  0.12 \, m$. 

Now let us assume for a moment that the mass $m$ is pushed up to $m \sim g {\cal M}/4\pi \sim g^2 M/4\pi$ by UV corrections.
Using Eq. (\ref{eq:efconstr})\ we find $g M/M_{pl} = \alpha^2 \pi/\sqrt{N}$. Combining this with the second equation in (\ref{eq:data_constraints}), we can solve for $\alpha$ to find $\alpha \sim 0.06$. Looking at Figure 1, there is still a small range of $g$ for which this barely sub-Planckian theory is viable, in the far left corner of the shaded region. 

Note that ${\cal M} \sim g M$ is at most a lower bound on the scale of new physics. In practice, new physics could appear up to the Planck scale, and thus push the technically natural scale for $m$ farther up still. In this case, the above argument shows that $\alpha$ is pushed up towards $1$. As with the Higgs mass, there is no clear barrier to new physics appearing up to the fundamental Planck scale, be this the 4d or 10d Planck scale. If $M$ is close to the Planck scale, $gM \lesssim 10^{15} \, {\rm GeV}$ (using the constraints on $g$ we will discuss below), so a higher fundamental scale indicates a more serious hierarchy problem beyond the existing need for a little hierarchy between $\mu\, ,  m$. This little hierarchy in itself is still a problem, albeit possibly a weaker one. 

In the end, controlling $\mu \, , m$ in the face of QFT and quantum gravity corrections requires unnatural fine-tunings or an explicit mechanism. In \S3\ we discuss one such mechanism, axion monodromy. This promotes the scalar masses to gauge field masses which are much less sensitive to UV corrections, whether from QFT or quantum gravity \cite{Kaloper:2016fbr} (this is not to say that the mass is completely unconstrained \cite{Kaloper:2016fbr,Reece:2018zvv}.) If the cutoff scale contributions 
$\sim g {\cal M}/4\pi$ to the masses $m, \mu$ are prohibited by a mechanism such as monodromy, the theory could be natural, UV safe and consistent with the data. We will show explicitly how this happens in hybrid monodromy in \S 3.

\subsubsection{Marginal couplings}

If we begin with the Stewart model  (\ref{eq:stewartmodel}) at tree level, we will induce not just mass terms but marginal quartic couplings $\frac{1}{4} \lambda \sigma^4$ and $\frac{1}{4} \lambda' \phi^4$.  The induced couplings $\lambda,\lambda'$ will be of order $g^4$ (times some factors we will discuss momentarily).  If we write down our EFT with couplings of this order from the outset, these couplings are technically natural.  If we choose $g$ sufficiently small, we will see that these couplings will remain subdominant and (\ref{eq:stewartmodel})\ is a good approximation to a model that is stable to quantum corrections in the regime that generates phenomenologically acceptable epochs of inflation and reheating. Thus, the marginal couplings by themselves are under control, in principle - if we were to ignore the relevant operators, the marginal couplings could be natural.

We estimate the size of the quantum corrections following the discussion of \cite{Coleman:1973jx}. The quantum corrections to  the quartic terms will include the term
\be
	{\cal L}_{radiative} \ni \frac{g^4}{64\pi^2} \ln\left(\frac{g^2 \phi^2}{\tilde {\cal M}^2}\right) \phi^4 \, .
	\label{cowa}
\ee
This set of terms arises from summing together the 1-PI irreducible diagrams renormalizing $\phi^4$ terms due to the virtual $\sigma$ modes.
Here $\tilde {\cal M}$ is the subtraction scale, which is in principle arbitrary. For example, it can be taken to be the mass scale of the $\sigma$ field during inflation, which is integrated out to generate the effective $\phi$ theory; or it might be the cutoff of the $2$-field 
hybrid model (since, practically, we do not expect that they are far apart). 

The log will give the largest contribution if the argument is small. This means, we will get the strongest correction to the scalar field potential for the smallest values of $\phi, \sigma$ during inflation -- i.e., near the exit. Since we already noted that the EFT during inflation will fail at the exit, needing a different EFT to describe reheating, we will here only require these corrections to remain under control during the early stages of inflation. To illustrate this quantitatively, let us take $g$ of order $10^{-3}$ and $\alpha \sim 0.1$, with the cutoff ${\cal M}\sim 10^{-2} M_{pl}$. In this case the log is of order unity; the loop factor $64\pi^2$ in the numerator gives a suppression factor of order $10^{-3}$, so this gives a correction to $\phi^4, \sigma^4$ of order $10^{-3} g^4$. Note that these values of $g,\alpha$ are within the allowed region in Fig. 1. 
 
The strongest constraint on the $\phi^4$ term is that its second derivative be subleading compared to the second derivative of the first two terms in (\ref{eq:inflation}).  As we discussed previously,  this means that the inflection point induced by the quartic corrections to (\ref{eq:inflation}) is far enough to allow the plateau to yield $\sim 50$ efolds of inflation. In our numerical example, $\ln(g^2\phi^2/\tilde {\cal M}^2) \sim 1$, and we find that this condition combined with 
Eqs. (\ref{eq:efconstr}) gives 
\be
g  \la 1.6 \times 10^{-3} \, .
\label{gbd}
\ee
This is the bound we incorporated preemptively in Fig. 1.  We see that this leaves a small region of parameter space that is consistent with both naturalness of the quartic couplings, and with sub-Planckian expectation values for $\phi,\sigma$. As noted, these scalars cannot be hugely sub-Planckian: in this range one or both are within an order of magnitude of $M_{pl}$.  Allowing for both scalars to have smaller ranges would require larger values of $g$. In turn this would require either finely tuned small quartic couplings for $\phi$ or some additional mechanism to suppress the self-coupling at all orders in the loop expansion. We of course assume that the theory is natural here, without large cancellations between regularized and bare terms in the loop expansion. 

Note that in \cite{Kaloper:2011jz}, we have discussed a very similar-looking problem with single field inflation, namely radiative corrections to the inflaton potential $V$. The point there was that the QFT corrections to $V(\phi)$ are automatically natural since
they are of the form $F(\eta) V$, where $F(\eta)$ is an analytic function of $\eta \sim \frac{M_{Pl}^2}{V} \partial_\phi^2 V$ at $\eta =0$, with
${\cal O}(1)$ expansion parameters. Thus during inflation, if the potential $V$ is chosen to be flat, $\eta \ll 1$ and loop corrections will remain small. In this model the new issue arises from the presence of the new field $\sigma$, and its new couplings to $\phi$. The Stewart model requires the biquadratic potential $\phi^2 \sigma^2$ to dominate the $\phi^2, \sigma^4$ terms. The bound (\ref{gbd}) ensures precisely that. 

Yet as we just stressed in the previous subsection, without some additional mechanism(s) to suppress the masses, and preserve a small hierarchy between them, it is \underline{not even possible} to keep the field ranges below the Planck scale without fine tuning. And even if we tune the field range for $M$, we still need $\alpha < 0.12$, which in turn implies  $g \sim 10^{-3}$, as designated in Figure 1. This shows just how fine tuned the theory must be -- baring hierarchy protection mechanisms.  One might hope that selecting a special scale in the log in Eq. (\ref{cowa}), which makes the log very small might help \cite{Lyth:1999sp}, however this would only make sense if the UV corrections are tamed.  Thus our discussion showcases just how desperately the model needs a mechanism in the UV to protect it from large quantum corrections to the masses.

\subsubsection{Irrelevant couplings}

Finally, let us consider the irrelevant operators. In addition to being a potential problem in QFT, these are also a portal for the effects of quantum gravity to come in. Let us first note that the latter are far more dangerous.  Basic dimensional considerations indicate that integrating out $\sigma$ during inflation will give corrections of the form\footnote{Aside from the overall prefactor $\sim 1/16\pi^2$ this term displays correct normalizations as
per NDA. We will however ignore these factors in this section, since our main purpose here is to outline the issue. Such additional numerical factors
may in fact be helpful.} 
\be
	\delta {\cal L}_p \sim \frac{g^{4 + 2p} \phi^{4 + 2p}}{{\cal M}^{2p}}
\ee
Taking the lower bound ${\cal M} \sim g M$ and applying the second equation in (\ref{eq:data_constraints}), we find
\be
	\delta {\cal L}_p \sim g^4 \left(\frac{g \alpha}{8\times 10^{-5}}\right)^{2p} \frac{\phi^{4 + 2p}}{M_{pl}^{2p}}
\ee
For $g \sim 10^{-3}$, $\alpha \sim 0.1$, and noting that we are ignoring phase space and symmetry factors, this is generally smaller by a factor of $g^4$ as compared to Planck-suppressed operators with ${\cal O}(1)$ coefficients. We will therefore focus on the latter but note that in general QFT contributions will also need to be suppressed.

Estimates such as the one above can be overly pessimistic. We know that in QFT, a series of operators which individually look dangerous, can sum up in the effective action such that the relevant effective potential remains flat. Essentially this can happen when the loop expansion is an alternating series, with operators of the form $\phi^{p + 4}$ having signs $(-1)^p$. As a result the sum total of all the operators which should be included in the EFT is merely a log correction to the leading term, as discussed in the previous section (see \cite{Kaloper:2011jz}). Thus the irrelevant operators then need not be a show-stopper, and indeed flattened potentials such as those discussed in \cite{Dong:2010in,DAmico:2017cda}\ depend on them.

However in the presence of the extra field $\sigma$, and with quantum gravity corrections having no known pattern, we will be maximally conservative, and instead outline the conditions which {\it guarantee} that operators irrelevant in the RG sense are also irrelevant in the sense of not contributing during inflation.  To this end, we will focus on the potentially the worst-behaved terms which are higher powers of the lighter field  $\phi$, normalized for convenience by the Planck scale $M_{pl}$. This is merely a matter of choice; a different normalization would yield apparently different numerical statements, but the contents would be exactly the same. 
So consider a coupling of the form
\be
	\delta V = \frac{\delta_p}{(p+4)!} \frac{\phi^{p+4}}{M_{pl}^p} \, .\label{eq:scalarirr}
\ee
If such corrections are too large, they behave as the inflaton mass term, shortening the width of the hybrid plateau of (\ref{potpotinfl}).
More specifically, the magnitude as well as derivatives of such operators during inflation need
to be smaller than the value and derivatives respectively of the tree-level potential in Eq. (\ref{eq:inflation}).  
Focusing on the second derivative, we find that
\be
	\delta_p \ll (p+2)! \, \alpha^{-p} \times 1.2 \times 10^{-13} \, .
	\label{irbound} 
\ee
Thus for $\alpha \sim 0.1$ we still need a mechanism which suppresses a finite set of irrelevant operators. This mechanism must suppress both QFT and quantum gravity corrections. 
To this end, a sub-Planckian axion, with or without monodromy, should be effective. We stress -- as we noted above -- that if there are cancellations between adjacent irrelevant operators in the EFT expansion of the effective potential, as in the case of large field models or Coleman-Weinberg theories where irrelevant operators 
comprise an alternating series, that could
help too. Our analysis of individual operators nevertheless shows that this may be easier to realize with sub-Planckian 
field ranges and parameters. As we have stressed all along, 
a mechanism which subverts  the UV sensitivity of the masses is beneficial, since it will also help with the irrelevant operators.

The upshot is that even though our model has sub-Planckian field ranges, the irrelevant operators are not
automatically guaranteed to be parametrically suppressed relative to the Planck scale.  The Planck-suppressed irrelevant operators of sufficiently low dimension might in fact interfere with slow-roll inflation unless their dimensionless couplings are kept sufficiently small.

\section{A pseudoscalar realization and its $4$-form dual}

Small field hybrid models of inflation, as exemplified by the Stewart model, face two serious issues:
\begin{itemize}
\item The scalar masses are UV sensitive within the confines of QFT;
\item There are Planck-suppressed irrelevant operators which require very small dimensionless coefficients.
\end{itemize} 
As we will review and develop here, both of these problems may be addressed by considering $\phi, \sigma$ as pseudoscalar axions dual to massive $4$-form field strengths \cite{Kaloper:2016fbr}. The masses $\mu, m$ are dual to the gauge theory masses, which are not UV sensitive; while corrections to the scalar potential are suppressed by additional powers of the ratio of the masses to the cutoff, ${m}/{\cal M}$, $\mu/{\cal M}$. These effects follow from the gauge symmetries of the model. In the duality frame described by the $4$-form field strengths, these are a pair of nonlinearly realized $U(1)$ gauge symmetries, with St\"uckelberg fields restoring the gauge symmetry of the mass term.  In the dual scalar theory the gauge group is discrete ${\mathbb Z}\times{\mathbb Z}$, with each factor acting on a scalar and on a discrete variable. These discrete variables are dual to $4$-form flux, labelling distinct branches of a multivalued potential \cite{Kaloper:2008fb,Kaloper:2011jz,Kaloper:2016fbr}, and act as a sort of discrete St\"uckelberg field. 

The symmetries are obscured on the scalar side by their nonlinear realization combined with gauge fixing, which follows from picking a specific branch where the scalar longitudinal modes of the massive $4$-form field strengths reside. Nevertheless the gauge redundancies remain operational in the full phase space of the theory, ensuring technical naturalness and protecting the scalar dynamics from the perils of quantum gravity. In the end, our goal here is to rewrite the Stewart model as an example of a simple gauge fixed EFT of massive $4$-forms, and demonstrate how the ills of the scalar theory may be healed by gauge symmetries. 

We find that for scalar theories satisfying the constraints outlined in \S2.2, with sub-Planckian field ranges,
the dual theory appears to be very strongly coupled: dimensionless coefficients of the leading
irrelevant operators, written as powers of the field strength, are pushed to be large. The 
duality map yields an NDA-like presentation of the pseudoscalar
action in terms of which the functions defining the potential appear to have large coefficients
in a Taylor series expansion. These apparent large couplings are the price we pay for our mechanism for controlling $m,\mu$,
while maintaining sub-Planckian field ranges in the face of constraints given by the data.

However, these coefficients may not be the proper measure of couplings, governing
the scattering amplitudes of asymptotic states. Absent mass terms, the $4$-forms are non-propagating; with masses, the propagating
modes are the longitudinal ones, which come multiplied by powers of the gauge field mass. Thus, the physical asymptotic states have couplings that are suppressed by ratios like $\mu/{\cal M}$,  $m/{\cal M}$. Wavefunction renormalizations (aka ``seizing'' \cite{Dimopoulos:2003iy}) can push
down the effective coupling further. The irrelevant operators induced by the leading ``large"
coupling are natural in the na\"ive sense -- that is, they have order ${\cal O}(\lesssim 1)$ dimensionless coefficients.

A complete exploration of naturalness and NDA for massive $p$-form gauge fields has not been carried out and we will not do so
here.  A UV completion of our $4$-form theory, would be an important way to explore the range of validity of our EFT.
We will outline the issues that we encounter and point the way to possible resolutions.

\subsection{Single field dual}

To illustrate the massive axion-massive $4$-form field strength duality, let us review the case of a single pseudoscalar field theory as in \cite{Kaloper:2016fbr,DAmico:2017cda}.  This procedure has been discussed previously in \cite{Julia:1979ur,Aurilia:1980xj,Dvali:2003br,Dvali:2005an}. We will review it in detail first,  adding some new 
interpretational comments which we hope are clarifying, and proceed to the two-field case.

We begin with the EFT action (in the $+---$ convention), 
\be
{\cal L} = \frac12 (\partial \phi)^2 - {\cal M}_*^4 V(\frac{\mu \phi}{{\cal M}_*^2}) \, .
\label{lagrone}
\ee
In \cite{Kaloper:2016fbr,DAmico:2017cda} we have normalized (\ref{lagrone}) using the rules of N\"aive Dimensional Analysis (NDA) \cite{Manohar:1983md,Gavela:2016bzc}. We have also chosen to canonically normalize the scalar kinetic term, and ignore the higher derivative terms 
which appear as irrelevant operators in the full effective action, but can be neglected on a subclass of backgrounds even when the argument of $V$ is large \cite{Kaloper:2016fbr,DAmico:2017cda}. Here it appears that we have inserted the mass scale $\mu$ (which is the scalar mass at small $\phi$) by hand. As we will see, on the dual gauge field side this is both the gauge field mass and the coupling of the longitudinal mode of said gauge field. This gives an avenue to tame the UV sensitivity of the masses, since gauge symmetry links them to couplings which are far less UV sensitive. So the point of this exercise is to take an apparently constrained Lagrangian and check if it arises from a natural one in the dual frame. 

The potential $V$ is the effective potential viewed as an asymptotic series in the argument.  As such we imagine calculating it to finite precision, truncating the theory to finitely many irrelevant operators.  In \cite{Kaloper:2016fbr,DAmico:2017cda}, we assumed each term in this series had order ${\cal O}(1)$ coefficients, so that in a sense the theory is strongly coupled. This allowed for the flattening of the scalar potential at large field ranges. In single-field, large-field-range inflation models we are pushed into this regime by the data: flatter potentials yield smaller tensor-scalar ratios and are therefore favored by the increasingly tightening bounds on $r$ by data. Thus in that analysis we ignored the additional source of small numbers in the effective action due to small dimensionless coefficients in $V$, which can be attained starting at strong coupling near the cutoff and then running the theory to lower scales. For this review we will
stick to the strongly-coupled theory, in order to focus on one set of issues at a time.

The central idea behind making the theory (\ref{lagrone}) natural is to promote $\phi$ to a {\it pseudoscalar axion} $\varphi$ with a periodicity $\varphi \to \varphi + f$, by defining $\mu \phi \equiv \mu \varphi + Q$. Here $Q = n e$, $n \in {\mathbb Z}$ is a discrete variable, and $e$ a discrete charge defining $\mu f = e$. The theory now has a discrete gauge symmetry ${\mathbb Z}$ acting as $\varphi \to \varphi + f$, $n \to n - 1$ \cite{Kaloper:2011jz,Kaloper:2016fbr}. The ${\mathbb Z}$ symmetry is spontaneously broken by picking a branch with the specific value of $Q$, which plays the role of the St\"uckelberg field for $\varphi$. Then $\phi$ ranges over a single branch of the multibranch potential. This branch structure, which was pointed out in discussions \cite{Witten:1978bc,Witten:1980sp}\ of the $\theta$-dependence of the vacuum for large-N gauge fields, is the basis of axion monodromy.

Given this discrete gauge symmetry, the form of (\ref{lagrone}) implies  that we should write the potential as a function of $\mu\varphi + Q$ rather than $\varphi + n f$.  Our strongest justification is that the dual $4$-form field strength theory can be written in terms of a (strongly coupled) natural theory, with a gauge field mass protected from UV corrections and irrelevant gauge-invariant operators suppressed by ${\cal M}_*$ with ${\cal O}(1)$ dimensionless coefficients.  We will thus maintain the attitude that at high energies (but below the cutoff), the fundamental theory is that of a massive $4$-form field strength, with mass $\mu$. Let us then review this
theory, and see how it leads to dual pseudoscalar theory of the form (\ref{lagrone}). 

The low energy EFT of the flux monodromy inflation model of \cite{Kaloper:2016fbr,DAmico:2017cda} is the most general massive abelian three-form gauge theory, including all the irrelevant operators allowed by gauge symmetry, and excluding ghosts.  The action is\footnote{Using the equations of motion to eliminate terms with $\p^k F$ factors, which encode spurionic massive ghosts at the cutoff.}
\ba \label{corrections}
&&{\cal L}^{\rm (full)}  = - \frac{1}{2 \cdot 4!} F_{\mu\nu\lambda\sigma}^2 - \frac{\mu^2}{12} (A _{\mu\nu\lambda} - h_{\mu\nu\lambda})^2   - \sum_{n>2} \frac{c_n}{n! {\cal M}_*^{2n-4}} {\tilde F}^{n}  \\
&&~~~~~ - \sum_{n > 1} \frac{c'_n }{2^n n! {\cal M}_*^{4n-4}} \mu^{2n} (A_{\mu\nu\lambda} -h_{\mu\nu\lambda})^{2n}   - \sum_{k\ge1, \, l\ge 1} \frac{c''_{k , l}}{2^k k! l! {\cal M}_*^{4k+2l-4}} \mu^{2k}
(A_{\mu\nu\lambda}-h_{\mu\nu\lambda})^{2k} {\tilde F}^{l}  \, , \nonumber
\ea
where $A$ is the gauge field 3-form, $F=dA$, ${\tilde F} = {}^*F = - \frac{1}{4!} \epsilon_{\mu\nu\lambda\sigma}  F^{\mu\nu\lambda\sigma}$, $b$ a two-form St\"uckelberg gauge field with field strength $h=db$, and $\mu$ plays the role of both the gauge field mass and the St\"uckelberg mode coupling. 
Gauge symmetry and the Goldstone Boson Equivalence Theorem (GBET) then assert that any power of $A$ not covered by a derivative must be multiplied by the same power of $\mu$ \cite{Kaloper:2016fbr}. Finally, 
${\cal M}_*$ is the strong coupling scale of the low energy EFT of the form sector, and 
${\cal M} = \sqrt{4\pi} {\cal M}_*$ is the cutoff. 
We have followed the canonical normalizations of the operators in (\ref{corrections}), having introduced
the appropriate combinatorial terms. This implies that in the strong coupling limit, near the cutoff ${\cal M}$, the coefficients $c_k$ are all ${\cal O}(1)$ -- within an order of magnitude. Clearly, our normalizations follow the NDA prescriptions of  \cite{Manohar:1983md,Gavela:2016bzc} for the $4$-form 
field strength $F$ and its associated $3$-form potential $A$, with the additional statement that all the couplings are set to their maximal value
where (\ref{corrections}) is still valid.

It is straightforward to dualize (\ref{corrections}) for small field values, near the vacuum $F=A=0$, where the action reduces to only quadratic terms. In this limit, we can proceed by 
treating the St\"uckelberg mode $h$ as a $3$-form whose exactness is enforced via a Lagrange multiplier 
$ \mu \varphi \epsilon^{\mu\nu\lambda\sigma} \partial_\mu h_{\nu\lambda\sigma}/6$, and also adding another Lagrange multiplier 
$Q \epsilon^{\mu\nu\lambda\sigma} (F_{\mu\nu\lambda\sigma} - 4\partial_\mu A_{\nu\lambda\sigma})/24$ which enforces $F=dA$. Then integrating out $h$ and $F$ yields the dual effective action
\be
{\cal L}_{quadratic}(\varphi, Q) = \frac12 (\partial \varphi)^2 
- \frac{(\mu \varphi +Q)^2}{2} + 
\frac16  \epsilon^{\mu\nu\lambda\sigma} Q \partial_\mu A_{\nu\lambda\sigma}\, ,
\label{quadr}
\ee
where the last term enforces $\partial Q = 0$. Here as we noted above $Q = N e$, where $e$ is the fundamental 4-form charge; $N \in \Z$; and $\phi \equiv \phi + f$ where $\mu f = e$ \cite{Kaloper:2011jz}. Thus clearly the canonically normalized field is $\phi = \varphi + Q/\mu$, as stated above.

It is useful to understand the dualization from the canonical point of view. Near the vacuum, integrating out the fields $F = dA$ and $h=db$, and absorbing $h$ into $A$ via a gauge transformation, leads to the conditions \cite{Kaloper:2016fbr,DAmico:2017cda}
\be
\mu \varphi +Q = - \frac{1}{4!} \epsilon_{\mu\nu\lambda\sigma} F^{\mu\nu\lambda\sigma} \, , 
~~~~~  - \epsilon_{\nu\lambda\sigma\mu} \partial^\mu \varphi = \mu A_{\nu\lambda\sigma} \, .
\label{dualizations}
\ee 
These are {\it canonical transformations}, since we are trading one `coordinate-momentum' conjugate pair for another. We treat these as IR dualities.  At the level of the quadratic terms, which dominate in the IR, this transformation preserves the form of the Hamiltonian of the theory, and so also the perturbative ground state. 

In the large field regime, for which the higher-order terms in (\ref{quadr}.\ref{corrections}) become important, 
the generalizations of the maps  (\ref{dualizations}) become more complicated. 
Nevertheless, we can outline the general procedure for how to proceed. 
The point is that both representations of the theory have
the same local degrees of freedom, and the same underlying dynamical symmetries. 
The $4$-form frame shows that the mass parameters are not UV sensitive (although they are not calculable in the EFT, and must be treated as inputs). This is precisely what we need to deal with the hierarchy problems of marginal operators in hybrid inflation \`a la Stewart. 

The procedure is conceptually the same as in the quadratic 
limit; we start with (\ref{corrections}), add the Lagrange multipliers 
$ \mu \varphi \epsilon^{\mu\nu\lambda\sigma} \partial_\mu h_{\nu\lambda\sigma}/6$ and 
$Q \epsilon^{\mu\nu\lambda\sigma} (F_{\mu\nu\lambda\sigma} - 4\partial_\mu A_{\nu\lambda\sigma})/24$, and go from there.
All we really need to do is treat $F$ and $h$ as independent variables in (\ref{corrections}) extended with the Lagrange multiplier 
terms, compute partial derivatives of the complete action with respect to them (and set $h =db$ at the end, to 
eliminate it using  gauge invariance of $A$), obtain (nonlinear) equations relating $\mu \varphi + Q$ and $\partial \varphi$ to $F$ and $A$, and invert them to determine the
coordinate change $A, F \rightarrow \partial \varphi, \, \mu \varphi + Q$. Since the $3$-form gauge potential $A_{\mu\nu\lambda}$ 
couples to a worldvolume of a membrane with charge $e$ -- and the volume of the gauge group is tuned to be compatible with the discrete membrane charge -- the associated $4$-form field strength $F_{\mu\nu\lambda\sigma} = 4 \partial_{[\mu} A_{\nu\lambda\sigma]}$ is then quantized in units of $e$. So ``integrating out" $F$ means summing over all the quantized values, which enforces the quantization of $Q$, also in the units of $e$.

A lesson we draw here is that the field $Q$ is central to this duality. As a discrete variable, it is dual to the transverse 
(non-propagating!) degrees of freedom of a $4$-form field strength. The field $\varphi$ is dual to the gauge mode, and acquires dynamics when the $4$-form theory is given a mass.

The resulting effective Lagrangian in terms of the scalar variable $\varphi$, which is valid up to the scale 
${\cal M} = \sqrt{4\pi} {\cal M}_*$, is
\ba \label{correctionsscalarnorm}
{\cal L} &=&  \frac{1}{2} (\partial_\mu \varphi)^2 - \frac{1}{2} (\mu \varphi + Q)^2  - \sum_{n>2} \tilde c_n 
\frac{(\mu\varphi + Q)^{n}}{ n! ({{\cal M}_*^2})^{n-2}} \, \nonumber \\
&& -  \sum_{n>1} \tilde c'_n \frac{(\partial_\mu \varphi)^{2n}  }{2^n n!({{\cal M}_*^2})^{2n-2}} - \sum_{k\ge 1, \, l \ge 1}  \tilde c''_{k,l}\frac{(\mu\varphi + Q)^{l} }{{2^k k! l! ({{\cal M}_*^2})^{2k+l-2}}}
(\partial_\mu \varphi)^{2k}  \, .
	\ea
where $\tilde c_n$ are in general complicated functions of $c_n$ from the dual form side. After the 
transformation we take them to be $\sim {\cal O}(1)$ by naturalness of the strong coupling expansion, unless they are prohibited by additional symmetries. The irrelevant operators in (\ref{correctionsscalarnorm}) can be activated
above the scale ${\cal M}_*$, without activating the UV degrees of freedom which only start to appear at scales 
above ${\cal M}$. 

As discussed in \cite{Kaloper:2016fbr,DAmico:2017cda}, the theory (\ref{correctionsscalarnorm}) supports two
distinct phases above the strong coupling scale, depending on the importance of the higher derivative operators
$(\partial \varphi)^{2k}$ with $k>1$. Here we will restrict our attention to only the phase where the higher derivatives are dynamically suppressed, and only the quadratic derivative terms play a significant role. In this limit (\ref{correctionsscalarnorm})\ can be formally resumed, yielding 
\be
{\cal L} = \frac12 {\cal Z}_{eff}(\frac{\mu \phi}{{\cal M}_*^2}) (\partial_\mu \phi)^2 - {{\cal M}^4}_*{\cal V}_{eff}(\frac{\mu \phi}{{\cal M}_*^2}) \label{eq:twod}\ ,
\ee
where ${\cal Z}_{eff}$ and ${\cal V}_{eff}$ are dimensionless functions, and the normalizations of the arguments reflect NDA and gauge symmetries. Here we defined $\phi = \varphi + Q/\mu$. Clearly, the field redefinition
$\sqrt{{\cal Z}_{eff}} \, d \phi \rightarrow d\phi$ canonically normalizes the scalar and puts (\ref{eq:twod}) in the form (\ref{lagrone}), rearranging the formal sum of the irrelevant operators activated above the strong coupling scale. 

Importantly, the duality transformation (\ref{dualizations}) changes the na\"ive dimension of an operator when the variables are exchanged. This occurs because we are replacing a field `coordinate' with a field `momentum' (i.e. derivatives of fields). 
Trading a field variable for a momentum lowers the dimension by two, while the converse raises the dimension by two. The kinetic term becomes a mass term and vice-versa. This is properly accounted by the NDA normalization in (\ref{lagrone}), which includes the powers of the mass with the field variables. Our procedure shows that this has a dual in which the mass appears as a gauge field mass, and the irrelevant operators (powers of the $4$-form) have a standard expansion in powers of $F$ divided by the strong coupling scale 
${\cal M}_*$. 

On the dual $4$-form side, the scalar field theory marginal couplings appear to be irrelevant as reflected by the change in normalizing dimensions by the powers of the cutoff.  As we stressed above, if the scalar self-interactions contain additional small parameters,  
the dual irrelevant operators will include small numbers which cannot be completely reproduced by the mass scales and wave function renormalizations.  Therefore in what follows for the sake of completeness we will treat leading longitudinal mode self-couplings as independent renormalized scalar theory couplings. In a natural theory, due to running, they will saturate at $\sim {\cal O}(1)$ near the cutoff. 
Well below the cutoff, at weak coupling, this may yield an additional reduction of the coefficients in the effective action as noted in \cite{Gavela:2016bzc}. Nevertheless, we will find that the requirement of naturalness of the theory and restriction to sub-Planckian scales and inflationary
field ranges pushes the theory out of this weak-coupling regime.

\subsection{Dual of a two field hybrid model}

We now turn to illustrating how to embed the family of two interacting scalar field theories that can support hybrid 
inflation, and contain the Stewart model as a limit into a dual theory of two interacting $4$-form field strengths. 
In this case, in addition to the complications arising due to nonlinear terms, the model also includes a term linear in $\sigma$. In principle, we could try to just shift the field until the linear term is absorbed away. However, $\sigma \sim 0$ gives the ``instantaneous" vacuum (in the Born-Oppenheimer sense) in the $\sigma$ sector at the beginning of inflation. 
As we noted above, in our version of hybrid inflation, the inflaton ($\phi$) mass goes up to the cutoff at the end of inflation; 
while the $\sigma$ mass drops below the cutoff as $\phi \to 0$, as seen in Eq. (\ref{sigmass}). Since 
we regard the pseudoscalar-$4$-form duality as an IR duality, with field and operator dimensions computed
from the IR fixed point, we must carefully specify the EFT and thus the scalar field value about which we perform the dualization.

In this section we will compute the dual form theory following these steps: 
\begin{itemize}
\item We will establish the dual correspondence of the canonical variables on the two sides, using NDA-normalized variables; in order to connect $\sigma$ and $\phi$ with their dual forms, we invoke the sector of the theory just at the end of inflation, but before
$\sigma$ relaxes to the true vacuum; here both scalars are light.
\item We will show that the scalar kinetic terms dualize to mass terms for the dual forms.
\item We will then identify the NDA-normalized non-linear couplings.
\item We will establish the mapping between the operators on the two dual sides in the weak coupling; we will identify the lower bound on the cutoff of the theory in terms of the dimensional parameters in the EFT. By weak coupling we mean that
dimensionless couplings for operators normalized by the cutoff are small, and the theory is in some sense close to Gaussian.
\end{itemize}
We are limiting this section to weak coupling for illustrative purposes. In \S3.3\ we will find that the bounds from naturalness and from observations, which we explored on the scalar side, appear to imply that the dual form theory must be in strong coupling during inflation. We will discuss the possible implications of this observation there.
The main result here will be a somewhat telegraphic walk through the duality transformation, skipping some of the explicit steps above; the reader can however  readily fill in the missing steps of the complete analysis.

Let us begin by establishing the canonical transformation between the scalar and $4$-form pictures, setting up the `dictionary' for 
transitioning from one side to the other. To simplify our formulae, we will use the dimensionless zero-form duals of the $4$-forms: 
\be
{\tt F} = - \frac{1}{4! {\cal M}_*^2} \epsilon_{\mu\nu\lambda\sigma}  F^{\mu\nu\lambda\sigma} \, , ~~~~~~~~~~
{\tt G} = - \frac{1}{4! {\cal M}_*^2} \epsilon_{\mu\nu\lambda\sigma}  G^{\mu\nu\lambda\sigma} \, , 
\label{notations}
\ee
Here $F = d A$, $G = d B$ locally; globally, the values $A, B$ in different charts of the cover of spacetime may be related by 2-form gauge transformations $A \to A - d\Lambda_A$, $B \to B - d\Lambda_B$ in the overlap between the charts.

The scale ${\cal M}_*$ in (\ref{notations}), included for dimensional reasons here, is the strong coupling of the scalar EFT (\ref{potpot}), as per NDA. We will link it to the theory's dimensional parameters below.  Note that since we are interested in dualizing a model of hybrid inflation for which higher-derivative terms do not contribute to the dynamics, we restrict our attention to only the terms which are quadratic in derivatives. 
Note that we will ignore terms of higher than quadratic order in $\p\phi, \p\sigma$.  In standard slow roll inflation, these terms are kept small by the dynamics \cite{DAmico:2017cda}.  As in that work there could be other regimes of the theory in which higher-derivative terms could also assist a slow-roll phase of the theory. A hybrid model with these higher-derivative terms activated would be an interesting topic for future work.

We expect that the dual scalar theory in general takes the form
\be
{\cal L} = \frac12 {\cal Z}(\frac{\mu \phi}{{\cal M}_*^2}, \frac{m \sigma}{{\cal M}_*^2}) (\partial \phi)^2  + \frac12 
\hat {\cal Z}(\frac{\mu \phi}{{\cal M}_*^2}, \frac{m \sigma}{{\cal M}_*^2}) (\partial \sigma)^2  - {\cal M}_*^4 \, {\cal V}(\frac{\mu \phi}{{\cal M}_*^2}, \frac{m \sigma}{{\cal M}_*^2})  \, .
\label{lagrtwofs}
\ee
As in the single-field case, we introduce the axions $\varphi$ and $\chi$, which are related to the hybrid inflation scalars $\phi$ and $\sigma$ via:
\be
\mu \phi = \mu \varphi + Q \, , ~~~~~~~~~ m \sigma = m \chi + P \, .
\label{scalarsforms}
\ee
This is to say that the discrete gauge symmetry ${\mathbb Z}\times {\mathbb Z}$ shifts $Q,\varphi$ and $P, \chi$ so that $\phi,\sigma$ remain unchanged. The compact scalars $\varphi, \chi$ are the duals to the longitudinal modes of the full massive form system: more precisely $d\varphi, d\chi$ are dual to the 3-form St\"uckelberg field strengths. They can be absorbed into the local fluctuations 
of the massive $3$-form potential via gauge fixing.

The functional form (\ref{lagrtwofs}), in which non-derivative couplings of $\phi, \sigma$ come multiplied by factors of
$\mu, m$,  and normalized by ${\cal M}_*$, is based on our experience with the single-field case. In the weak coupling limit we study here, this form is justified in that it produces a natural theory of $4$-forms, in the sense of NDA. It should be possible to justify this combination of mass parameters and scalar field values entirely within the scalar  frame by utilizing 
discrete gauge invariances of 
the model, with the explicit discrete St\"uckelberg fields $P,Q$, but we leave this for future work.

To formally dualize the scalar theory, we have found it useful to employ the dimensionless variables
\be
\Phi = \frac{\mu \varphi +Q}{{\cal M}_*^2}\, , ~~~~~~~~~~ 
{\cal X} = \frac{m\chi +P}{{\cal} {\cal M}_*^2} \, , 
\label{dimensiscals}
\ee
and then rewrite the potential in (\ref{lagrtwofs}) in terms of them.  We call these ``NDA-normalized variables" as the potential ${\cal V}$
and the kinetic functions ${\cal Z}$, ${\hat {\cal Z}}$, can be written in terms of them. Furthermore, imagine for the moment that 
we can approximate ${\cal Z}, \hat {\cal Z} \simeq 1$. Adding and subtracting the
Lagrange multiplier terms  $\Phi {\tt F} + {\cal X} {\tt G}$,  the  ``chimera" Lagrangian that ensues is
\be
{\cal L} = \frac12 (\partial \varphi)^2  + \frac12 (\partial \chi)^2 + \frac{ \mu \varphi}{4!} \epsilon_{\mu\nu\lambda\sigma}  F^{\mu\nu\lambda\sigma} + \frac{ m \chi}{4!} \epsilon_{\mu\nu\lambda\sigma}  G^{\mu\nu\lambda\sigma} -
{\cal M}_*^4 \Bigl( {\cal V}(\Phi, {\cal X}) - \Phi~ {\tt F} - {\cal X} ~ {\tt G} \Bigr) \, ,
\label{lagrtwofsnew}
\ee
which, after integrating the scalar-$4$-form bilinears by parts, becomes 
\be
{\cal L} = \frac12 (\partial \varphi)^2  + \frac12 (\partial \chi)^2 - \frac{ \mu}{3!} \epsilon_{\mu\nu\lambda\sigma}  \, \partial_\mu \varphi  \, A^{\nu\lambda\sigma} + \frac{m }{3!} \epsilon_{\mu\nu\lambda\sigma} \, \partial_\mu \chi  \, B^{\nu\lambda\sigma} -
{\cal M}_*^4 \Bigl( {\cal V}(\Phi, {\cal X}) - \Phi~ {\tt F} - {\cal X} ~ {\tt G} \Bigr) \, .
\label{lagrtwofsnew2}
\ee
The first four terms dualize to mass terms for $A$, $B$: we can complete the squares for the scalar derivatives and integrate them out, with the remaining terms being precisely $\mu^2 A^2$ and $m^2 B^2$. This part is straightforward because this contribution to the Lagrangian is bilinear at weak coupling. 

What remains is to dualize the effective potential, and replace the variables $\Phi$ and ${\cal X}$ with ${\tt F}$ and ${\tt G}$ defined in (\ref{notations}). We can treat the last four terms in (\ref{lagrtwofsnew2}) independently from the rest because $\Phi, \XX$ are combinations of the discrete St\"uckelberg fields $Q,P$, and so can be varied independently of $\varphi,\sigma$.
The $4$-form dual of the final term in brackets in (\ref{lagrtwofsnew2})\ is the Legendre transform of the effective potential. In practice, we integrate out the fields  $\Phi, {\cal X}$ to replace them with ${\tt F}, {\tt G}$, which means inverting \cite{Dvali:2003br,Dvali:2005an}
\be
{\tt F} = \partial_\Phi {\cal V} \, , ~~~~~~~~~~ {\tt G} = \partial_{\cal X} {\cal V} \, , 
\label{cantrfleg}
\ee
and substituting $\Phi = \Phi({\tt F}, {\tt G})$, ${\cal X} = {\cal X}({\tt F}, {\tt G})$ into $K =  \Phi~ {\tt F} + {\cal X} ~ {\tt G} - {\cal V}(\Phi, {\cal X})$. In doing this, we bear in mind that during inflation the scalar $\sigma$ is much heavier than $\phi$ -- in fact it may be heavier than the cutoff ${\cal M} \sim g M$. 
Since it is changing very slowly, with initial value $\sigma \simeq 0$, the field $\sigma$ remains displaced from its true minimum at
$\sqrt{2} M$ for a period after inflation ends, when $\sigma$ is much lighter than during inflation. 
Thus we can pick the transient value of $\sigma \simeq 0$ as a ``pivot" to dualize the scalar theory at it. This means that we should dualize $\sigma$ around zero during inflation and right after inflation, and around $\sqrt{2} M$ at the very late stages after inflation when much of reheating takes place, when $\sigma$ oscillates around the true minimum. The region near $\sigma = \sqrt{2} M$, about which the theory reheats, and the ``plateau" at $\sigma \sim 0$ describing inflation and its end prior to reheating, are different 
phases; they are best treated as distinct EFTs as we already explained in \S2. In the latter, $\sigma$ remains heavier than $\phi$. This is true even for small $\phi$ when we choose $m \gg \mu$, as we do if we wish $\phi$ to be the inflaton. Near $\sigma = \sqrt{2} M$, $\phi$ becomes the heavy field with a mass at or near the cutoff. These phases must in general be connected inside a UV completion.

We compute the duality transformation for the case that ${\cal V}$ is has the functional form (\ref{potpot}): in terms of $\Phi$, ${\cal X}$, this is:
\be
{\cal V} = \frac{1}{2} \Phi^2 + \frac{1}{2} \bigl({\cal X} - \gamma\bigr)^2 +  \frac{{\bar g}^2}{4} \Phi^2 {\cal X}^2 + \frac{\bar \lambda}{4} \Bigl({\cal X}^2 - \delta^2\Bigr)^2 + \frac{{\bar \lambda}'}{4} \Phi^4 + \ldots \, , 
\label{potpotbig}
\ee
where the rescaled couplings are:
\be
\gamma = \sqrt{2} \frac{m M}{{\cal M}_*^2} \, , ~~ \delta = \frac{m\tilde M}{{\cal M}_*^2} \, , ~~ \bar \lambda = \lambda \, \bigl(\frac{{\cal M}_*}{m}\bigr)^4 \, , ~~ \bar g^2  = g^2 \, \bigl(\frac{{\cal M}_*^2}{ m  \mu} \bigr)^2 \, , ~~ \bar \lambda' = \lambda' \, \bigl(\frac{{\cal M}_*}{ \mu}\bigr)^4 \, . ~
\label{couplings}
\ee
These are the couplings that appear in the $4$-form dual; we have not written out the operators that would be
irrelevant in the scalar frame. We expect that a proper formulation of NDA for this theory will involve potentials and coefficients
of $(\p\phi)^2$, $(\p\sigma)^2$ that are functions of $\Phi$, ${\cal X}$, ${\bar g}^2/16\pi^2$, $\lambda/16\pi^2$, $\lambda'/16\pi^2$.
The appearance of couplings with factors of $1/16\pi^2$ generally arises in effective actions due to phase space factors in loop integrals \cite{Gavela:2016bzc}.

In this section, to provide an explicit example, we will work in the approximation $\bar \lambda, \delta, \bar \lambda' \ll \bg^2 \la 1$, and drop terms proportional to $\blam,\blam',\delta$. This limit is consistent with the functional form of the Stewart model. Note that this
``weak coupling" assumption is stronger than the assumption that $\lambda, \delta,\lambda' \ll g^2 \la 1$; as we will see when
we impose consistency with the constraints of \S2, data pushes ${\bar g}^2$ to be large. Yet it can remain the domain
of strong coupling below the cutoff of the NDA-normalized action. We will retain only the lowest order irrelevant operators on the dual form side, as they are duals of the marginal operators on the scalar side. We will neglect writing out explicitly the higher dimension irrelevant operators here; we will however need to discuss them in \S3.3, when we go to comparatively large $\phi$ in order to describe the epoch of inflation that imprints on the CMB.

To solve Eqs. (\ref{cantrfleg}), we solve the nonlinear equations  one at a time, and perform the expansion of the algebraic inversions of (\ref{cantrfleg}) in Taylor series in the couplings (\ref{couplings}).  This means, formally, that we take the ``instantaneous vacuum" to be controlled by the root of the nonlinear equations (which, to any finite order in couplings, are polynomials), dominated by the linear terms. The solutions remain perturbative in the coupling constants, and thus consistent with the EFT description. Using (\ref{potpotbig}) in (\ref{cantrfleg}), inverting, and expanding in the couplings, we find:
\begin{eqnarray}
\Phi & = & {\tt F} \Bigl(1 - \frac{\bar g^2}{2} {\cal X}^2 \Bigr) + {\cal O}(\bar \lambda, \bar \lambda', \bar g^4) \nonumber\\
{\cal X} & = & \Bigl( {\tt G} + \gamma \Bigr) \Bigl(1 - \frac{\bar g^2}{2} \Phi^2 \Bigr) + {\cal O}(\bar \lambda, \bar \lambda', \bar g^4) \, .
\label{trafos}
\end{eqnarray}
Since we want solutions which are perturbative in the coupling, we can use each of these equations to ${\cal O}(1)$ to replace the terms to ${\cal O}(\bar g^2)$ in the other, and find the correct
answer to the order $\bar g^2$. Thus, the inversion formulas are 
\ba
\Phi &=& {\tt F} \Bigl(1 - \frac{\bar g^2}{2} \bigl( {\tt G} + \gamma\bigr)^2 \Bigr) + {\cal O}(\bar \lambda, \bar \lambda', \bar g^4) \, , \nonumber \\
{\cal X} &=& \Bigl( {\tt G} + \gamma \Bigr) \Bigl(1 - \frac{\bar g^2}{2} {\tt F}^2 \Bigr) + {\cal O}(\bar \lambda, \bar \lambda', \bar g^4) \, .
\label{invertransf}
\ea
Substituting these into  $K =  \Phi~ {\tt F} + {\cal X} ~ {\tt G} - {\cal V}(\Phi, {\cal X})$ and using (\ref{potpotbig}) finally
yields
\be
K = \frac12 {\tt F}^2 + \frac12 \Bigl( {\tt G} + \gamma \Bigr)^2 - \frac12 \gamma^2 - \frac{\bar g^2}{4} {\tt F}^2  \Bigl( {\tt G} + \gamma \Bigr)^2 + {\cal O}(\bar \lambda, \bar \lambda', \bar g^4) \, .
\label{legendreK}
\ee
Using (\ref{notations}), 
the total Lagrangian in the dual variables and at weak coupling is therefore
\be
{\cal L} = {\cal M}_*^4 \Bigl\{ \frac12 {\tt F}^2 + \frac12 \Bigl( {\tt G} + \gamma \Bigr)^2 - \frac12 \gamma^2 - \frac{\bar g^2}{4} {\tt F}^2  \Bigl( {\tt G} + \gamma \Bigr)^2  \Bigr\} + \frac{\mu^2}{12} A_{\nu\lambda\sigma}^2 
+ \frac{ m^2}{12} B_{\nu\lambda\sigma}^2 + {\cal O}(\bar \lambda, \bar \lambda', \bar g^4) \, .
\label{lagrfivehybrid}
\ee
where again $F = dA$, and $G = dB$, and ${\tt F}, {\tt G}$ are defined in terms of 
$F, G$ in Eqs. (\ref{notations}). 

Note that from the first of Eqs. (\ref{invertransf}), the vacuum of the theory ${\cal X} = 0$ maps to ${\tt G} + \gamma = 0 + {\cal O}(\bar g^2)$. This means that ${\tt G}$ does not correctly describe the fluctuations of the dual form sector $B, G$ around this vacuum. Instead the correct canonical variable to use is
\be
{\cal G}  = {\tt G} + \gamma \, .
\label{ggg}
\ee
This constant shift of the magnetic dual of the $4$-form field strength implies a shift of the electric $4$-form
${\cal G}_{\mu\nu\lambda\sigma} = {\cal M}_*^2 \, \epsilon_{\mu\nu\lambda\sigma} \, {\cal G}$ by
\be
{\cal G}_{\mu\nu\lambda\sigma} = {G}_{\mu\nu\lambda\sigma} + \gamma {\cal M}_*^2 \, \epsilon_{\mu\nu\lambda\sigma} \, ,
\ee
or in the form notation, ${\cal G} = {\tt G} + \gamma {\cal M}_*^2 \, \Omega_4$, where $\Omega_4$ is the space-time volume $4$-form. If we write $\cG = d\cB$, then the $3$-form potentials are are related by
\be
{\cal B}_{\mu\nu\lambda} = B_{\mu\nu\lambda} + h_{\mu\nu\lambda}
\ee
Here $h$ is defined locally, by integrating
\be
d h = \gamma {\cal M}_*^2 \, \Omega_4 \, . 
\ee
to yield 
\be
	h = \gamma {\cal M}_*^2 \, t \, \Omega_3\, , \label{eq:forminitcond}
\ee
where $\Omega_3$ is the volume form of the constant comoving time hypersurfaces. 

With all of this, the renormalizable and leading irrelevant terms in the dual theory at weak coupling 
around the transient vacuum $\sigma \simeq 0$ are:
\ba
{\cal L} &=& - \frac{1}{2 \cdot 4!} F_{\mu\nu\lambda\sigma}^2 - \frac{1}{2 \cdot 4!} {\cal G}_{\mu\nu\lambda\sigma}^2 
- \frac12 \gamma^2 {\cal M}_*^4 - \frac{\bar g^2}{4 \cdot (4!)^2 {\cal M}_*^4}  F_{\mu\nu\lambda\sigma}^2  {\cal G}_{\mu\nu\lambda\sigma}^2 \nonumber \\
&& + \frac{\mu^2}{12} A_{\nu\lambda\sigma}^2 
+ \frac{ m^2}{12} \Bigl({\cal B}_{\mu\nu\lambda} - h_{\mu\nu\lambda} \Bigr)^2 + {\cal O}(\bar \lambda, \bar \lambda', \bar g^4) \, .
\label{lagrfivehybridreal}
\ea
This is the dual form EFT of the Stewart model limit (\ref{eq:stewartmodel}) of hybrid inflation in the regime $\sigma \simeq 0$ and
$g \phi < m$ (in the case that $\blam, \blam' \ll {\bar g}^2 \ll 1$). This region of field space, which is reached right after inflation ends, is readily accessible to simple analytical tools to construct the dual. 
Note that this regime can be readily extrapolated down to $\phi = 0$ as long as $\sigma \simeq 0$. 
However the extrapolation down to $\sigma = \sqrt{2} M$ requires a different EFT, because at $\sigma = \sqrt{2} M$ the
$\phi$ field gets Higgsed by the $\sigma$ {\it vev}, having the mass $\mu^2_{eff} \simeq g^2 M^2 \sim {\cal M}^2$ (\ref{phimass}). 
Next we wish to discuss the limit of the theory that extends beyond this regime and supports inflation -- which requires $g \phi \gg m$. As we have stated, and will see in detail shortly, that forces us to go beyond weak coupling and the action (\ref{lagrfivehybridreal}). 

Note, that the
situation here is analogous to what we encountered in the single field case. The scalar and massive $4$-form duals are related simply
in the weak coupling limit, with the scalar being the light and weakly coupled longitudinal mode of the massive $4$-form field strength.
Yet to produce inflation that fits the data the theory must be pushed into the strong coupling, beyond the simple perturbative
picture, where the connection between the scalar and the form fields becomes very nonlinear. Nevertheless, we take the attitude
that the existence of the dual form picture suffices for our purpose since it is a `home' for the gauge symmetries that protect 
the scalar mass. Once that is in force, to study inflation one can work on the scalar side alone, and evolve the theory to strong coupling.

Before continuing with the foray into the inflationary regime, let us clarify the role of the $\gamma$-dependent terms in Eq. (\ref{lagrfivehybridreal}). These are the positive vacuum energy $\half\gamma^2 \MM_*^4$, and the $3$-form $h$ inside the mass term for $\cB$. The former is the initial value of the vacuum energy driving inflation. In the dual scalar picture, this comes from the $\sigma$ mass term near $\sigma = 0$, and assumes that any additional contribution to the vacuum energy is negligible during inflation. The additional cosmological contributions which we neglected here would be the vacuum energy in the true vacuum, and neglecting them corresponds to choosing the vacuum energy in the true vacuum 
to be very small. This is just the cosmological constant problem, which is notoriously difficult to address in any EFT. This appearance of the vacuum energy as a constant in our EFT points to the fact that this appears to be fine tuned. This fine tuning could be addressed via saltatory variation of the cosmological constant by nucleation of membranes charged under the $4$-forms, following \cite{Brown:1987dd,Brown:1988kg,Bousso:2000xa}; regions with a different initial vacuum energy but the same couplings otherwise will either inflate forever, or collapse too soon.

Absent the mass term here, the only way to relax the initial vacuum energy towards zero is via membrane nucleation. In our theory, the mass term allows this vacuum energy to {\it also} be relaxed by the slow rolling of the longitudinal mode of the massive form. Here the 3-form $h$ in the mass term ensures that the pivot point $\cG,\cB = 0$ correctly describes the onset of inflation in terms of variables with canonical kinetic terms. It appears as a source term in the equation of motion $d^* \cG + m^2 \cB = m^2 h$, driving the longitudinal mode to evolve so that the vacuum energy is lowered.

\subsection{Scales and couplings for hybrid monodromy EFT}

We now turn to considering pseudoscalar axions in the parameter regime controlled by the dual requirements that the theory
have sub-Planckian field ranges and be consistent with current data. Our aim is to write an effective action which respects 
NDA normalizations, in which we are assured that the mass parameters $\mu, m$ are UV-insensitive, and the irrelevant
operators are suppressed.

The first thing we might try is to perform the duality map outlined in the previous section. The masses $\mu, m$
map to $4$-form masses which we know are UV-insensitive; while past experience in the more extreme
case of large-field inflation is that the irrelevant operators do not spoil slow-roll inflation 
\cite{Kaloper:2008fb,Kaloper:2011jz,Kaloper:2014zba,Kaloper:2016fbr,DAmico:2017cda}.
From the discussion above, one would expect that a natural effective action of $4$-forms following 
\cite{Gavela:2016bzc}\ would be a function of $F/\MM_*^2$, $G/\MM_*^2$, $m$, $\mu$, ${\bar g}^2/16\pi^2$, 
$\blam/16\pi^2$, $\blam'/16\pi^2$, with the factors of $1/16\pi^2$ arising from loop factors. 

As we noted, however, and in full analogy with the single large field theories, in the parameter regime of the Stewart model pointed to by the data, 
the duality map is no longer controlled by the leading linear map $F \sim \mu \phi$, ${\cal G} \sim m \sigma$.
We can see from Eq. (\ref{trafos}) that this would require $\bg \Phi < \sqrt{2}$; we will see that this does not hold 
during the epoch of inflation that imprints on the observed CMB. 

Inspired by the map at weak coupling, however, we could hope that there is an effective action 
in the scalar frame which satisfies some version of NDA. At weak coupling, the
duality map indicates that $\mu, m$ are not renormalized (though we currently lack an argument directly in the
scalar frame).  It is then natural to suppose that the effective
action would be a function of $\MM_*$, $\Phi$, ${\cal X}$, ${\bar g}^2/16\pi^2$, $\blam/16\pi^2$, $\blam'/16\pi^2$
with ${\cal O}(1)$ coefficients, as the weak-coupling duality points in this direction. 

However, we will see that in the regime allowed by the data, $\bg^2/16\pi^2$ is large, and when the field ranges are lowered by an
order of magnitude below the Planck scale and more, perturbativity of the theory appears to be in jeopardy. A concern is that the effective action is strongly coupled and out of control. On the other hand, we will see that with $\alpha {\cal O}(\la 1)$ we can still meet the data
and suppress irrelevant operators enough while keeping $\bar g^2/16\pi^2$ barely within the range of EFT. One possible take from this is 
that identifying $\bg^2$ defined in Eq. (\ref{couplings}) as the coupling may be overly na\"ive. 
On the scalar side, $\Phi$, ${\cal X}$ are of course not canonically normalized, and moving to canonical variables demonstrates that the natural couplings are $g^2,\lambda,\lambda'$. On the $4$-form side, at weak coupling, 
the massive $4$-forms $F$ only propagate in the presence of a mass term for the gauge fields: the two-point functions $\vev{{}^* F {}^*F} \propto \mu$, $\vev{{}^* G {}^*G} \propto m$ up to contact terms, so that even $\frac{{\bar g}^2}{16\pi^2} \ga 1$ can induce ${\cal O}(1)$ values of ${\bar \lambda}$, ${\bar \lambda}'$ via quantum corrections. The large effective $\bar g^2$, which is required by combining naturalness and data, can come about as a result of nontrivial nonlinear mixings of various irrelevant operators on backgrounds with large form fluxes, that simulate the inflationary vacuum energy. In other words, the scale of the biquadratic operator
$\sim g^2 \phi^2 \sigma^2$ on the scalar side is not set by a single large irrelevant operator in the dual theory but
is a combination of many irrelevant operators of dimension higher than eight, which add together
enhancing the effective coupling. In the end, it is an open question how to implement naturalness and NDA for massive 4-forms or their duals\footnote{And for that matter, NDA for massive vector fields.}.  We leave this for future work.

In the remainder of this section we develop the above points in detail, and highlight possible paths towards ensuring hybrid inflation
makes sense both phenomenologically and as a natural QFT safe from quantum gravity.

\subsubsection{Identification of NDA parameters}

We open with identifying the range of parameters $\MM_*$, $\bg^2$, $\blam$, $\blam'$ in the potential (\ref{potpotbig}). We dub this the NDA potential for the scalar fields. We will conjecture that the discrete gauge symmetries together with dimensional analysis demand that $\phi$, $\sigma$ appear in the form $\Phi$, ${\cal X}$. We would further demand that the couplings of the theory are consistent with the values induced by quantum corrections; that is, that the model is self-consistently technically natural, or even simply natural.
At present we do not know how this would work in practice, just as we do not completely understand NDA for massive $4$-forms.
Should the duality hold, there are nonrenormalization theorems in the scalar theory that we have not derived that will constrain
the effective action. A simple guess, by rough analogy with \cite{Gavela:2016bzc}, would be that if all couplings were ${\cal O}(16\pi^2)$, our theory would be technically natural if the effective potential could be writen as $V(\Phi, {\cal X})$ with ${\cal O}(1)$ couplings; if $\bg^2/16\pi^2  \ll 1$, our theory could be natural for $\blam/16\pi^2$, $\blam'/16\pi^2$ to be small; and for $\bg^2/16\pi^2 \gg 1$, the theory is strongly coupled and completely out of control. 

Before turning to the couplings, our first question is the choice of the cutoff scale $\MM$ and the strong 
coupling scale ${\cal M}_* = \MM/\sqrt{4\pi}$. In a top-down theory these would of course be fundamental quantities. Here we are taking a bottom-up approach, asking what values of the cutoff give an action such that the dual form theory takes an NDA-like form. 
We thus identify $\MM_*^4$ as the scale of the energy density over which the effective potential varies, and in 
particular the scale that drives inflation:
\be
{\cal M}_*^2 \ga \sqrt{2} mM \, .
\label{strongcoupling}
\ee
The scale of the energy density driving inflation is thus $V \sim m^2 M^2 \la {\cal M}_*^4/2 = {\cal M}^4/{32\pi^2}$. Given the subtleties
we are about to discuss in identifying expansion coefficients in NDA for this theory, there may be some wiggle room here. We will simply
adopt our straightforward definition of $\MM_*$ and see where it gets us. Note that this definition of $\MM_*$ guarantees that the
field range of ${\cal X}$ between the end of inflation and reheating is ${\cal O}(1)$.

Next, we wish to bound $\bg^2$. Equation (\ref{eq:efolds}), giving the number of efolds of inflation in terms of the field displacement $\phi$, gives (after a few lines of algebra):
\be
N = \alpha^2 \pi^2 \frac{\bar g^2}{16\pi^2} \Phi^2 \, .
\label{NDEfolds}
\ee
Demanding now that $\alpha < 1$ and $N \ga 50$ immediately shows that to have inflation we must start with $\bar g^2 \Phi^2$ which is initially at least as big as
\be
\bar g^2 \Phi^2 \ga \frac{16 N}{\alpha^2} \ga 800 \, .
\label{Phiinit}
\ee
If this were the only constraint, we could support sufficient inflation in a regime where the coupling ${\bar g}^2/16\pi^2$ is small. At weak coupling, $\Phi \sim 4\pi$ is a unitarity bound. This maps to the statement $F/\MM_*^2 < 4\pi$, $F/\MM^2 < 1$. If this bound on $\Phi$ extends past the regime that the duality map can be constructed perturbatively in powers of the fields, we can ask when the unitarity limit saturates the inequality in Eq. (\ref{Phiinit}). This occurs when
$\bg^2 \geq 5$, or $\bar g^2/16\pi^2 \ga 0.032$. Taking $\bg^2/16\pi^2$ to be the right parameter for an NDA analysis of the scalar action, the effective action should still be under perturbative control even if the tree-level coupling looks strong.

A further constraint on ${\bar g}$ comes from imposing the observed scalar power which combine with the number of efolds leads to
Eq. (\ref{eq:data_constraints}). We begin with 
\be
	\frac{g^2}{16\pi^2} = \frac{\bg^2}{16\pi^2}\left(\frac{m\mu}{\MM_*^2}\right)^2 \, ,
\ee
and employ our bound (\ref{strongcoupling}). Furthermore, we will assume a hierarchy $\mu = \eps m$; $\eps = 1/8$ corresponds to the bound from $M$ being sub-Planckian.
Finally, using Eq. (\ref{strongcoupling}), and then Eq. (\ref{eq:data_constraints})\ to write $m/M$ in terms of $\alpha, g$ we find:
The resulting lower bound on $\bg^2$ is:                                                                                                                                                                                                                                                                                                                                                                                                                                                                                                                                                                                                                                                                                                                                                                                                                                                                                                                                                                                                                                                                                                                                                                                                                                                                                                                                                                                                                                                                                                                                                                                                                                                                                                                                                                                                                                                                                                                                                                                                                                                                                                                                                                                                                                                                                                                                                                                                                                                                                                                                                                                                                                                                                                                                                                                                                                                                                                                                                                                                                                                                                                                                                                                                                                                                                                                                                                                                                                                                                                                                                                                                                                                                                                                                                                                                                                                                                                                                                                                                                                                                                                                                                                                                                                                                                                                                                                                                                                                                                                                                                                                                                                                                                                                                                                                                                                                                                                                                              
\be
	\frac{\bg^2}{16\pi^2} \gtrsim  \frac{6\times 10^{-8}}{\eps^2 \alpha^6 g^2} \, .
	\label{eq:bargconstraint}
\ee
If we were to set $\eps = 1/8$, as required for $M < M_{pl}$, and choose $g \sim 1.6\times10^{-3}$, the maximum value allowed for the Stewart model to be technically natural, then $\bar g^2/16 \pi^2 \la 1$ would require $\alpha >  1$. This can be satisfied with a just barely
super-Planckian field displacement. The coupling becomes strong rather quickly if we lower $\alpha$ while fixing $g^2, \eps$. There is thus a tension between keeping the scalar theory technically natural and sub-Planckian, and our criterion that the NDA couplings be ${\cal O}(\lesssim 1)$.

If we further input the technically natural scalings $\lambda,\lambda' \sim g^4$, we find
\begin{eqnarray}
	\bar{\lambda} & \sim & g^4 \left(\frac{\MM_*}{m}\right)^4 \sim \bg^4 \left(\frac{\mu}{\MM_*^4}\right) \, , \nonumber\\
	{\bar\lambda}' & \sim & g^4 \left(\frac{\MM_*}{\mu}\right)^4 \sim \bg^4 \left(\frac{m}{\MM_*}\right)^4 \, .
\end{eqnarray}
Thus $\bg^2 \gtrsim 1$ is still consistent with ${\cal O}(\lesssim 1)$ couplings $\blam,\blam'$.
If $\mu \sim 0.3 g \MM_*$, then $\blam'/16\pi^2 \sim {\cal O}(1)$. For $m = 8\mu \sim 2 g \MM_*$, $\blam \sim 2\times 10^{-4}$. 

We have computed these values of $\blam,\blam'$ from technically natural values of the couplings in the scalar theory, as discussed in \S2. In terms of our NDA variables they indicate that a consistent application of the principles behind NDA -- that couplings are of the same order as their quantum corrections -- allows for some complicated structure of the action in the variables $\Phi, {\cal X}$. This should not be surprising as these variables are not canonically normalized, and their correlation functions will scale as positive powers of $m,\mu$ relative to those for $\phi, \sigma$. But it is these variables which naturally map to the dual $4$-forms.

Finally, we can ask what happens if operators such as $\Phi^{4 + k}$ appear in the action with ${\cal O}(1)$ coefficients. From the
discussion above, this could be overly pessimistic from the point of view of technical naturalness, but it is expected that a UV
completion that includes quantum gravity will enhance irrelevant operators from their technically natural values. We are assuming that said completion will still give couplings that are functions of $\mu\phi$, $m\sigma$ weighted by powers of $\MM_*$ or $M_{pl}$. Let us
consider the former scale, to be maximally conservative within our set of conjectures. Then we find
\be
	\delta {\cal L}_p \sim \frac{c_p }{(p+4)!}\frac{\mu^{4+p} \phi^{4+p}}{\MM_*^{4 + 2p}} \, . \label{eq:ndairrel}
\ee
Comparing this to Eq. (\ref{eq:scalarirr}), we find
\be
	\delta_p \sim c_p \frac{M_{pl}^p \, \mu^{p+4}}{ \MM_*^{2p+4}} \, .
\ee
If we saturate our bound Eq. (\ref{strongcoupling}), let $\mu = \eps m$, and impose the constraints Eq. (\ref{eq:data_constraints}),
we find:
\be
	\delta_p \sim \left(\frac{\alpha g \eps}{1.1 \times 10^{-4}}\right)^p \left(\frac{\alpha^3 g^2 \eps^2}{3.1 \times 10^{-3}}\right)^2 c_p \, .
\ee
This is consistent with Eq. (\ref{irbound})\ if 
\be
	\left(\frac{\alpha^2 g \eps}{1.1 \times 10^{-4}}\right)^p \left(\frac{\alpha^3 g^2 \eps^2}{3.1 \times 10^{-3}}\right)^2 c_p
	\ll (p+2)! \, \times 1.2 \times 10^{-13} \, .
	\label{irelbound} 
\ee
If we adopt $g \sim 1.6 \times 10^{-3}$ and take $\epsilon \simeq 0.1$, this bound translates into
\be
	\left( 1.45 \times \alpha^2 \right)^p \, \alpha^6 \, c_p
	\ll (p+2)! \, \times 2 \times 10^{-3} \, ,
	\label{irelbound2} 
\ee
and is readily achieved for all $p \ge 1$ even when $c_p \sim 1$ as long as $\alpha \le 0.7$. In this regime, using Eq. (\ref{eq:bargconstraint}), $\bar g^2/16\pi^2 \simeq 1.5/\alpha^6$, we find that for 
$\alpha \simeq 0.7$ the coupling becomes $\bg^2/16\pi^2 \simeq {\cal O}(10)$. This demonstrates that there is some wiggle room with the numbers, where we can either adjust the cutoff ${\cal M}$ down by a factor of a few, or take a slightly larger dimensionless coefficient of the biquadratic operator to reduce 
the effective coupling $\bar g$ while maintaining control over irrelevant operators. 
So for our theory to hang together, all we need is to keep $c_p \sim 1$ even though $\bg^2/16\pi^2$ is large, consistent with some notion of naturalness. From our discussion above, meeting this requirement does not seem out of reach. However we see rather dramatically how naturalness and data press the theory against Planck scale. 

\subsubsection{Comments on the dual massive $4$-form theory}

Our conjectures for writing an action for $\phi, \sigma$ consistent with NDA and the discrete gauge symmetries was inspired by
the $4$-form dual, in a regime for which the duality transformation can be computed and simply understood. 
However, the regime of the Stewart model supporting inflation consistent with data and sub-Planckian scalar fields is well out of this regime. 
We can see from the transformation (\ref{trafos}) that our iterative procedure for constructing the duality map begins to break down when $\bg \Phi > \sqrt{2}$.  Indeed, (\ref{Phiinit}) shows that 50 efolds before the end of inflation -- the epoch during which
inflaton fluctuations imprint on the CMB -- we are well out of this range for  $\alpha \le 20$. This value of $\alpha$ would defeat the original purpose of hybrid inflation, and we will set it aside.

Thus, we do not have complete control of the $4$-form dual when the parameters of the Stewart model are consistent
with the data and when the field values are in the range which is relevant for the CMB. Nevertheless the scalar 
theory exists in this regime, and maps to the dual $4$-form theory cleanly in the small field limit 
where we understand the duality. Thus, we could follow the theory in either frame as we increase the couplings and field values. The small field regime attained as inflation ends then serves as  the anchor for this `theory flow': as long as we are sufficiently close to $\phi, \sigma \sim 0$, then $\bg \Phi$ will become arbitrarily small and the duality transformation is under control. In this regime we can derive insights into the pseudoscalar dual. Let us then discuss aspects of the duality map in this regime. 

When the duality pertains, the effective action 
\ba
{\cal L} &=& - \frac{1}{2 \cdot 4!} F_{\mu\nu\lambda\sigma}^2 - \frac{1}{2 \cdot 4!} {\cal G}_{\mu\nu\lambda\sigma}^2 
- \frac12 \gamma^2 {\cal M}_*^4 - \frac{c_1}{4 \cdot (4!)^2 {\cal M}_*^4}  F_{\mu\nu\lambda\sigma}^2  {\cal G}_{\mu\nu\lambda\sigma}^2 \nonumber \\
&+& \frac{ c_2}{ (4!)^2 {\cal M}_*^4} \Bigl( F_{\mu\nu\lambda\sigma}^2 \Bigr)^2 + \frac{ c_3}{(4!)^2 {\cal M}_*^4} \Bigl({\cal G}_{\mu\nu\lambda\sigma}^2 \Bigr)^2  + \frac{\mu^2}{12} A_{\nu\lambda\sigma}^2 
+ \frac{ m^2}{12} \Bigl({\cal B}_{\mu\nu\lambda} - h_{\mu\nu\lambda} \Bigr)^2 + \ldots \, . ~~~
\label{stewdualgen}
\ea
where we have included all dimension-8 operators consistent with the symmetries, should capture the pseudoscalar dynamics well.
In this action, $c_1 \sim {\bg}^2$; $c_2 \sim \blam$, $c_3 \sim \blam'$. The results which we find in this regime will be corrected 
at larger coupling and larger field values, but as long as the theory remains below the unitarity bound and the fluctuating light longitudinal 
modes have couplings suppressed by $\mu/{\cal M}, m/{\cal M}$, the qualitative insights gained by this analysis may continue to the phenomenologically
relevant case of large $\bg \Phi$.

First, the parameters $\mu, m$ are  UV-insensitive and are at most logarithmically divergent. This follows from the arguments given in \cite{Kaloper:2016fbr}\ for the single field case. We suspect that we can run this argument directly in the pseudoscalar dual, using the nonlinearly realized $\ZZ\times\ZZ$ discrete gauge symmetry. Realizing this would give further support for 
maintaining this feature of the theory in the large $\bg \Phi$ regime.

Secondly, we can treat $M$ as a derived quantity. Recall this appears as a source term for $\sigma$ in the dual theory, and
sets the field range in $\sigma$ between inflation and the end of reheating.
Once we fix the cutoff ${\cal M}$, then if the bound (\ref{strongcoupling}), is saturated, $M = \MM_*^2/\sqrt{2}m$ is given by a see-saw formula. In particular it is a derived quantity.  In weak coupling,  (\ref{strongcoupling}) can be translated into a bound on the maximal $4$-form flux above which our EFT (\ref{lagrfivehybridreal}) breaks down. Since ${\cal M}^2 \sim 4\pi \, G_{max} \sim 4\pi \, {\cal N}_{max} \, e$, then Eq. (\ref{strongcoupling})\ implies:
\be
M \la \frac{{\cal M}^2}{\sqrt{2} m} \sim {\cal N}_{max} \, \frac{e}{\sqrt{2} m} \sim   {\cal N}_{max} \, \frac{f}{\sqrt{2}}  \, ,
\label{Meq}
\ee
where $e$ is the fundamental charge of membranes charged under ${\cal G}$ and $f = e/m$ is  the period of the dual axion $\sigma$. 
The upshot is that as a derived quantity, the scale $M$ can exceed the cutoff ${\cal M}$ without violating naturalness. Note here that $ {\cal N}_{max} f$ represents the field range of $\sigma$ during inflation; we are simply saying that field {\it ranges}\ need not be bounded by the cutoff, a fact we already understand for axions without monodromy (see for example \cite{Kaloper:2015jcz}). 

Finally, while it is tempting to identify $c_1 \sim \bg^2$ as a coupling in terms of which we would write an effective action following the rules of NDA, the actual story is more complex. The essential point is that $F, {\cal G}$ only propagate because of the gauge field mass. Their two-point functions scale as $\mu^2, m^2$ respectively. This can be seen directly by consistency with the duality transformations $F \sim\mu\phi$, ${\cal G} \sim m \sigma$. Alternatively, we can simply compute the propagator for $F,{\cal G}$. Given the propagator presented in \cite{Kaloper:2016fbr}\ for a massive $3$-form potential 
\be
	\langle A_{\mu\nu\lambda}(p) A_{\mu'\nu'\lambda'}(-p) \rangle  = 
		 \epsilon_{\mu\nu\lambda\rho}\epsilon_{\mu'\nu'\lambda'\rho'}\left( \frac{\frac{\xi}{2}\eta^{\rho\rho'}}{p^2 - \frac{\xi \mu^2}{2}} + \frac{\left(1 - \frac{\xi}{2}\right) p^{\rho}p^{\rho'}}{(p^2 - \mu^2)(p^2 - \frac{\xi \mu^2}{2})}\right)\, ,
\ee
where $\xi$ is a gauge-fixing parameter, the propagator for $F$ is
\be
	\langle {}^*F(p) {}^*F(-p) \rangle = C \left(1 + \frac{\mu^2}{p^2 - \mu^2}\right) \, ,
\ee
where $C$ is a dimensionless constant comprised of symmetry factors. The first term in parentheses is a contact term; removing this by taking appropriate account of operator mixing, we find ${}^*F$ behaves $\mu$ times a scalar, consistent with the duality. Because
propagators scale with the $4$-form masses, calculations that are perturbative in $c_k$ will come with additional powers of $m/\MM_*$, $\mu/\MM_*$. For example, let us ask whether $c_1 \gg c_3$ is consistent with quantum corrections.  The $F$ propagators yield 
\be
	c_3 \sim \bg^4  \left(\frac{\mu}{{\cal M}_*}\right)^4 \sim \left(\frac{ g {\cal M}_*}{m}\right)^4 \, .
	\ee
In this regime the scaling should come as no surprise -- when ${\cal G} \sim m \sigma$, this is compatible with the technically
natural value $\lambda \sim g^4$. Our conclusion is that in a proper treatment of the effective action for $4$-forms, extending the
principles of Na\"ive Dimensional Analysis, $c_1 \sim \bg^2$ will appear in combination with powers of $\mu, m$ so that the natural
coupling is some ${\hat g}^2 \ll \bg^2$; and that a large value of $\bg$ is still compatible with a sensible effective action for which 
$c_{k > 1}$ can be ${\cal O}(\lesssim 1)$.

Can we extend this observation to higher-dimension irrelevant operators in this regime? Consider the operators
\be
	\delta {\cal L}_k = c_k \frac{F^{2k}}{M_*^{4k-4}} \, , \label{eq:Firrel}
\ee
which are the most important for the dynamics in a phase ${\cal G} \sim m\sigma \sim 0$, $F\neq 0$.
At weak coupling, these terms are dual to a series of operators with leading terms (\ref{eq:scalarirr}), with $p = 2k - 4$. These operators will
be generated after integrating out the heavy $\sigma$ field, or on the dual side, from considering 
diagrams with ${\cal G}$ internal lines\footnote{Which contribute to the virtual momentum transfer due to the 
propagating longitudinal mode.}. The simplest diagram generating (\ref{eq:Firrel}) involves $k$ vertex insertions
$\propto \bar g^2 F^2 {\cal G}^2/{\cal M}_*^4$ in a ${\cal G}$ loop. This gives
$\delta {\cal L}_k \simeq {\bar g^{2k }}\frac{F^{2k}}{{\cal M}_*^{4k}} \langle ({\cal G}^2)^k \rangle$, so that after the loop
integral we find, using $\langle ({\cal G}^2)^k \rangle \simeq {\cal M}_*^4 (m/{\cal M})^{2k}$ (note the cutoff in the numerator as opposed to the
strong coupling scale),
\be
\delta {\cal L}_k \simeq  {\cal M}_*^4 \bar g^{2k}\frac{F^{2k}}{{\cal M}_*^{4k}}  \Bigl( \frac{m^2}{4\pi {\cal M_*}^2} \Bigr)^{k} 
\simeq {\cal M}_*^4 \Bigl(\frac{\bar g^{2} m^2}{4 \pi {\cal M}_*^2}\Bigr)^k\frac{F^{2k}}{{\cal M}_*^{4k}} 
\simeq {\cal M}_*^4 \Bigl(\frac{g^{2} {\cal M}_*^2}{4 \pi \mu^2}\Bigr)^k\frac{F^{2k}}{{\cal M}_*^{4k}} \, .
\ee
The last equality looks dangerous due to the to appearance of the small mass $\mu$ is the denominator. First, we note that
that the limit $\mu \rightarrow 0$ is not a problem since $F \sim \mu \phi/{\cal M}_*^2$ so $\mu$ precisely cancels if we
move back to the pseudoscalar frame. Furthermore, if $4\pi \mu^2 \gtrsim g^2 \MM_*^2$, it appears that 
$c_k \sim {\cal O}(1)$ is consistent even as $\bg$ is increased.

More precisely, if we substitute $\mu \sim m/8$ the overall dimensionless factor multiplying the NDA-normalized part of the operator $\delta {\cal L}_k$ becomes, using Eq. (\ref{strongcoupling}), and substituting (\ref{eq:data_constraints}) for $M/m$, 
\be
c_k \sim \Bigl(\frac{1.6 \times g {\cal M}_*}{m}\Bigr)^{2k} \ga  \Bigl(2.54 \times \frac{g^2 \sqrt{2} M}{m}\Bigr)^{k} \simeq \Bigl(\frac{0.2}{\alpha}\Bigr)^{3k} \, .
\label{girreldual}
\ee
Remarkably this shows that the irrelevant operator contributions generated by integrating out $\sigma$ remain safely small for
even sub-Planckian field displacements $\alpha \ga 0.2$, which as we noted we need to enforce to suppress their corrections to the 
inflationary plateau of (\ref{potpotinfl}). Further note that for $k=2$, the loop-induced term is $c_2 F^4/{\cal M}_*^4$, i.e. just the radiative correction  to $\bar \lambda'$. We therefore find that $\delta \bar \lambda' \sim (0.2/{\alpha})^6$, which, again, is under control unless $\alpha$ is too small.\footnote{Here of course we are looking near $\phi = 0$ where the duality makes sense;
$\alpha$ has meaning in the dual pseudoscalar theory as the range of $\phi$ in field space covered by the last $50$ efolds of inflation,
a range over which the duality map becomes complicated.}
If this holds, then the resulting dual operators are precisely of the form (\ref{eq:ndairrel}), with coefficients
$c_p \sim 1$, and these are subleading during inflation for $\alpha \lesssim 0.5$. 
Data pushes the theory towards the dangerous Planckian region, but there is still a consistent sub-Planckian
regime in which the irrelevant operators are under control.

A rigorous understanding of the $4$-form theory in the inflating regime is a matter of future work. Here we simply note that
it is plausible to have a well-defined theory with irrelevant operators built from powers of $F,$ ${\cal G}$, $\mu A$, and $m ({\cal B} - h)$  having ${\cal O}(\lesssim 1)$ dimensionless parameters when normalized via $\MM_*$, even when the leading coefficient 
$c_1 \sim \bg^2 \gg 1$. In this case we would need to show in this theory that the phenomenologically relevant phase
of inflation would occur for $F < \MM^2 = 4\pi \MM_*^2$, where we expect our effective theory to be well-defined. 

To conclude this section, we have seen that the theory (\ref{stewdualgen}) and its special limit (\ref{lagrfivehybridreal}) whose strong coupling limit can realize hybrid inflation\footnote{Again, such a theory can generate inflation if for a given cutoff ${\cal M}_*$ the masses obey $\mu \ll m \ll {\cal M}_*$ and one of the form field strengths develops an initial CP-breaking flux  on the  background, controlled by $M_{cr} \sim {\cal M}_*^2/m$.} are 
theories of two massive $U(1)$ gauge theories of $4$-form field strengths in the unitary gauge, 
with $4$-form kinetic mixings mediated by irrelevant operators. The mixing coefficients are controlled by natural parameters of ${\cal O}(1)$. In this form the theory is strongly coupled but natural: the
unbroken gauge symmetries of (\ref{stewdualgen}) will protect the selection of the masses and couplings in (\ref{stewdualgen}) from UV effects in QFT. However, since the scales are all sub-Planckian, and the symmetries are gauged, the theory will be safe from quantum gravity corrections as well.
The mass terms cannot receive large UV corrections since they are also couplings of the longitudinal modes, and are protected by gauge redundancies. We cannot predict what the values of these masses are from within the EFT itself. But once chosen, the gauge symmetries of the theory protect them from QFT and quantum gravity corrections, as in \cite{Kaloper:2016fbr,DAmico:2017cda}. Furthermore, while the masses are small relative to the cutoff, they will not be {\it too}\ small; they can be a few ($< 10$) orders of magnitude below the Planck scale. Thus the theory should be able to pass the bounds which one might find using arguments based
on the weak gravity conjecture, and from other technical lamppost-based bounds coming from recent developments in string phenomenology. 

\subsection{Discussion}

We have outlined how hybrid inflation might be made UV complete via dualizing it
to a theory of two massive $4$-form field strengths/$3$-form potentials. This 
UV completion contains gauge symmetries which explain the suppression of potentially dangerous operators 
which can adversely affect inflation \cite{Kaloper:2016fbr,DAmico:2017cda}. In a top-down approach to deriving hybrid inflation, this procedure would be reversed.

Indeed, imagine that in some UV-complete theory such as string theory, some of the
higher rank forms yield massive $4$-forms after compactification, with masses which are much smaller than
the UV cutoff of the EFT of the $4$-form systems. We believe that such constructions would be conceptually 
similar to the  previous approaches \cite{Silverstein:2008sg,McAllister:2008hb,Dong:2010in,Marchesano:2014mla,Hebecker:2014eua,Blumenhagen:2014gta,Grimm:2014vva,Dudas:2014pva,Ibanez:2014swa} where the main focus was on realizing the conditions for single large field inflation. In the present case they should involve multiple massive $4$-forms below the cutoff. The EFTs of interest arise after integrating out the stabilized heavier fields, the KK states and the heavy string modes. The cutoff ${\cal M}$ demarcates this EFT from the  full theory with those
additional degrees of freedom, and could be viewed as scale where ignoring the lightest of the modes, which were integrated out to define the low energy EFT, would start yielding problems with unitarity.  The irrelevant operators in (\ref{stewdualgen}) arise as the corrections generated by the virtual heavy modes  as well as loops of the virtual light modes kept in the EFT. In the minimal case, these are the longitudinal modes of the massive $4$-forms and the matter degrees of freedom, that the longitudinal modes decay into at the end of inflation. Perhaps the simplest manner in which these models can be realized is to imagine a theory with coupled
$p$-forms, which includes higher-derivative corrections suppressed by a cutoff, and where after dimensional reduction the emergent $4$-forms mix with pseudoscalar axions. These can be set to become longitudinal modes by a gauge fixing; after dualization, the higher-derivative operators are the potential for the longitudinal mode.

The longitudinal modes remain light because the $4$-form/$3$-form potential gauge symmetries: continuous compact $U(1)$ and the discrete shift, $A \rightarrow A + da$, ${\cal B} \to {\cal B} + db$. These ensure that the dangerous corrections to the mass terms are absent. This extends to the quantum gravity corrections as well, which cannot break gauge symmetries. 
So as long as the operators in (\ref{stewdualgen})
are below the cutoff ${\cal M}$, the theory has a weak coupling expansion where a minor tuning of parameters
realizes the regime which supports hybrid inflation. 

This is manifest once one reverses the steps which led from (\ref{eq:stewartmodel}) to (\ref{stewdualgen}). Indeed, inverting the steps in \S3.2 will map the
mass terms and irrelevant operators in (\ref{stewdualgen}) precisely on the potential (\ref{eq:stewartmodel}). The mass terms are naturally small by gauge symmetry, whereas the marginal operator couplings in 
(\ref{eq:stewartmodel}) are rendered small because they are controlled by the ratio of masses and the cutoff, and
$m_{A,B} \ll {\cal M}$. Finally, the scale $M$ in (\ref{eq:stewartmodel}) appears to be larger than the cutoff 
because it is see-sawed by the mass: $M \sim {\cal  M}^2/m$ as we discussed above. This can explain the origin of the Stewart limit of hybrid inflation naturally. 

In this paper we have focused on realizing a model of hybrid inflation in the Stewart limit, controlled by 
only relevant and marginal operators. Our observation that the $4$-form theory needs to be strongly coupled suggests
that we consider models in which other higher-dimension operators play a significant role. In particular, it is possible that  phenomenologically interesting and natural low-scale hybrid models could be realized with flattened potentials, following \cite{Silverstein:2008sg,McAllister:2008hb,Dong:2010in,DAmico:2017cda}. It would be interesting
to explore such more general models of hybrid inflation. 

\section{Summary}

Many inflationary models have been severely constrained by the observations in the past  decade or  so. Specifically the improving bounds on the spectral index  and on the tensor-scalar ratio have put pressure on the large field inflation models,  which are arguably the  simplest candidates for natural EFTs of inflation. These models are quite
predictive as  well, because they  must occur  at high scales  to yield viable inflationary  evolution, where their structure becomes sensitive to quantum gravity corrections. Thus the tightening constraints on large field models  might be taken to imply that the  prospects of learning something about quantum gravity from inflation have diminished. Moreover, by placing increasingly tighter constraints on large field models, observations might  appear to favor more exotic, unnatural, proposals for  inflation, or even more radical approaches to early universe cosmology. 

Our results here suggest that these allusions are not a foregone conclusion yet. The pressures from observational bounds are significantly relieved by reducing the scale of  inflation. This can be done in multifield inflation models,
and it occurs in hybrid inflation with two non-degenerate  fields, when the post-inflationary vacuum manifold is not
degenerate. In this case, the tensor-scalar ratio is almost unobservably small, and the spectrum of perturbations is
safely red, with the spectral index $n_S$ between $0.97$  and $0.975$, which is still in agreement with the bounds.
Further, the EFT of this variant of hybrid inflation is technically natural, and if it is realized as a dual for a theory with two massive $4$-forms, which might be realized as an IR limit of string compactifications, it may also be protected
from quantum gravity corrections although it involves almost Planckian field displacements. The UV safety of the theory is not a generic feature of all hybrid inflation proposals, as we have seen in detail. Yet it may arise in some constructions such as those which we outline here. However, remarkably, even in these cases the natural EFTs are still close to Planck scale. This, in our view, is quite interesting, since it keeps the possibility open that inflation, while being a viable EFT, might still be sensitive to some subleading corrections from quantum gravity, which while small 
might compete with the UV field theory effects.

\section*{Acknowledgements}

We would like to thank G. D'Amico and A. Westphal for useful discussions. N.K. would like to thank the CERN Department of Theoretical Physics and KITP, UCSB for kind hospitality in the course of this work. N.K. is supported in part by DOE Grant DE-SC0009999, and in part in part by the National Science Foundation under Grant No. NSF PHY-1748958 (to KITP). A.L. is supported in part by DOE grant DE-SC0009987. J.S. is supported 
in part by DOE Grants DE-SC0009999 and DE-0019081.

\bibliography{hybridrefs.bib}

\providecommand{\href}[2]{#2}\begingroup\raggedright\begin{thebibliography}{10}

\bibitem{Linde:1983gd}
A.~D. Linde, ``{Chaotic Inflation},''
\href{http://dx.doi.org/10.1016/0370-2693(83)90837-7}{{\em Phys. Lett.}
  {\bfseries B129} (1983) 177--181}.

\bibitem{Linde:1987yb}
A.~D. Linde, ``{Chaotic inflaton with constrained fields},''
\href{http://dx.doi.org/10.1016/0370-2693(88)90006-8}{{\em Phys. Lett.}
  {\bfseries B202} (1988) 194}.

\bibitem{Freese:1990rb}
K.~Freese, J.~A. Frieman, and A.~V. Olinto, ``{Natural inflation with pseudo -
  Nambu-Goldstone bosons},''
\href{http://dx.doi.org/10.1103/PhysRevLett.65.3233}{{\em Phys. Rev. Lett.}
  {\bfseries 65} (1990) 3233--3236}.

\bibitem{Adams:1992bn}
F.~C. Adams, J.~R. Bond, K.~Freese, J.~A. Frieman, and A.~V. Olinto, ``{Natural
  inflation: Particle physics models, power law spectra for large scale
  structure, and constraints from COBE},''
  \href{http://dx.doi.org/10.1103/PhysRevD.47.426}{{\em Phys. Rev.} {\bfseries
  D47} (1993) 426--455},
\href{http://arxiv.org/abs/hep-ph/9207245}{{\ttfamily arXiv:hep-ph/9207245
  [hep-ph]}}.

\bibitem{Kim:2004rp}
J.~E. Kim, H.~P. Nilles, and M.~Peloso, ``{Completing natural inflation},''
  \href{http://dx.doi.org/10.1088/1475-7516/2005/01/005}{{\em JCAP} {\bfseries
  0501} (2005) 005},
\href{http://arxiv.org/abs/hep-ph/0409138}{{\ttfamily arXiv:hep-ph/0409138
  [hep-ph]}}.

\bibitem{Silverstein:2008sg}
E.~Silverstein and A.~Westphal, ``{Monodromy in the CMB: Gravity Waves and
  String Inflation},'' \href{http://dx.doi.org/10.1103/PhysRevD.78.106003}{{\em
  Phys. Rev.} {\bfseries D78} (2008) 106003},
\href{http://arxiv.org/abs/0803.3085}{{\ttfamily arXiv:0803.3085 [hep-th]}}.

\bibitem{Kaloper:2008qs}
N.~Kaloper and L.~Sorbo, ``{Where in the String Landscape is Quintessence},''
  \href{http://dx.doi.org/10.1103/PhysRevD.79.043528}{{\em Phys.Rev.}
  {\bfseries D79} (2009) 043528},
\href{http://arxiv.org/abs/0810.5346}{{\ttfamily arXiv:0810.5346 [hep-th]}}.

\bibitem{Kaloper:2008fb}
N.~Kaloper and L.~Sorbo, ``{A Natural Framework for Chaotic Inflation},''
  \href{http://dx.doi.org/10.1103/PhysRevLett.102.121301}{{\em Phys. Rev.
  Lett.} {\bfseries 102} (2009) 121301},
\href{http://arxiv.org/abs/0811.1989}{{\ttfamily arXiv:0811.1989 [hep-th]}}.

\bibitem{McAllister:2008hb}
L.~McAllister, E.~Silverstein, and A.~Westphal, ``{Gravity Waves and Linear
  Inflation from Axion Monodromy},''
  \href{http://dx.doi.org/10.1103/PhysRevD.82.046003}{{\em Phys.Rev.}
  {\bfseries D82} (2010) 046003},
\href{http://arxiv.org/abs/0808.0706}{{\ttfamily arXiv:0808.0706 [hep-th]}}.

\bibitem{Berg:2009tg}
M.~Berg, E.~Pajer, and S.~Sjors, ``{Dante's Inferno},''
  \href{http://dx.doi.org/10.1103/PhysRevD.81.103535}{{\em Phys.Rev.}
  {\bfseries D81} (2010) 103535},
\href{http://arxiv.org/abs/0912.1341}{{\ttfamily arXiv:0912.1341 [hep-th]}}.

\bibitem{Dong:2010in}
X.~Dong, B.~Horn, E.~Silverstein, and A.~Westphal, ``{Simple exercises to
  flatten your potential},''
  \href{http://dx.doi.org/10.1103/PhysRevD.84.026011}{{\em Phys.Rev.}
  {\bfseries D84} (2011) 026011},
\href{http://arxiv.org/abs/1011.4521}{{\ttfamily arXiv:1011.4521 [hep-th]}}.

\bibitem{Kaloper:2011jz}
N.~Kaloper, A.~Lawrence, and L.~Sorbo, ``{An Ignoble Approach to Large Field
  Inflation},'' \href{http://dx.doi.org/10.1088/1475-7516/2011/03/023}{{\em
  JCAP} {\bfseries 1103} (2011) 023},
\href{http://arxiv.org/abs/1101.0026}{{\ttfamily arXiv:1101.0026 [hep-th]}}.

\bibitem{DAmico:2012khf}
G.~D'Amico, R.~Gobbetti, M.~Schillo, and M.~Kleban, ``{Inflation from Flux
  Cascades},'' \href{http://dx.doi.org/10.1016/j.physletb.2013.07.050}{{\em
  Phys. Lett.} {\bfseries B725} (2013) 218--222},
\href{http://arxiv.org/abs/1211.3416}{{\ttfamily arXiv:1211.3416 [hep-th]}}.

\bibitem{Kaloper:2014zba}
N.~Kaloper and A.~Lawrence, ``{Natural chaotic inflation and ultraviolet
  sensitivity},'' \href{http://dx.doi.org/10.1103/PhysRevD.90.023506}{{\em
  Phys. Rev.} {\bfseries D90} no.~2, (2014) 023506},
\href{http://arxiv.org/abs/1404.2912}{{\ttfamily arXiv:1404.2912 [hep-th]}}.

\bibitem{McAllister:2014mpa}
L.~McAllister, E.~Silverstein, A.~Westphal, and T.~Wrase, ``{The Powers of
  Monodromy},'' \href{http://dx.doi.org/10.1007/JHEP09(2014)123}{{\em JHEP}
  {\bfseries 09} (2014) 123},
\href{http://arxiv.org/abs/1405.3652}{{\ttfamily arXiv:1405.3652 [hep-th]}}.

\bibitem{Kaloper:2016fbr}
N.~Kaloper and A.~Lawrence, ``{London equation for monodromy inflation},''
  \href{http://dx.doi.org/10.1103/PhysRevD.95.063526}{{\em Phys. Rev.}
  {\bfseries D95} no.~6, (2017) 063526},
\href{http://arxiv.org/abs/1607.06105}{{\ttfamily arXiv:1607.06105 [hep-th]}}.

\bibitem{DAmico:2017cda}
G.~D'Amico, N.~Kaloper, and A.~Lawrence, ``{Monodromy Inflation in the Strong
  Coupling Regime of the Effective Field Theory},''
  \href{http://dx.doi.org/10.1103/PhysRevLett.121.091301}{{\em Phys. Rev.
  Lett.} {\bfseries 121} no.~9, (2018) 091301},
\href{http://arxiv.org/abs/1709.07014}{{\ttfamily arXiv:1709.07014 [hep-th]}}.

\bibitem{Nomura:2017ehb}
Y.~Nomura, T.~Watari, and M.~Yamazaki, ``{Pure Natural Inflation},''
\href{http://arxiv.org/abs/1706.08522}{{\ttfamily arXiv:1706.08522 [hep-ph]}}.

\bibitem{DAmico:2018mnx}
G.~D'Amico, N.~Kaloper, and A.~Lawrence, ``{Strongly Coupled Quintessence},''
  \href{http://dx.doi.org/10.1103/PhysRevD.100.103504}{{\em Phys. Rev.}
  {\bfseries D100} no.~10, (2019) 103504},
\href{http://arxiv.org/abs/1809.05109}{{\ttfamily arXiv:1809.05109 [hep-th]}}.

\bibitem{Baumann:2010ys}
D.~Baumann and D.~Green, ``{Desensitizing Inflation from the Planck Scale},''
  \href{http://dx.doi.org/10.1007/JHEP09(2010)057}{{\em JHEP} {\bfseries 09}
  (2010) 057},
\href{http://arxiv.org/abs/1004.3801}{{\ttfamily arXiv:1004.3801 [hep-th]}}.

\bibitem{Linde:1993cn}
A.~D. Linde, ``{Hybrid inflation},''
  \href{http://dx.doi.org/10.1103/PhysRevD.49.748}{{\em Phys. Rev.} {\bfseries
  D49} (1994) 748--754},
\href{http://arxiv.org/abs/astro-ph/9307002}{{\ttfamily arXiv:astro-ph/9307002
  [astro-ph]}}.

\bibitem{Copeland:1994vg}
E.~J. Copeland, A.~R. Liddle, D.~H. Lyth, E.~D. Stewart, and D.~Wands, ``{False
  vacuum inflation with Einstein gravity},''
  \href{http://dx.doi.org/10.1103/PhysRevD.49.6410}{{\em Phys. Rev.} {\bfseries
  D49} (1994) 6410--6433},
\href{http://arxiv.org/abs/astro-ph/9401011}{{\ttfamily arXiv:astro-ph/9401011
  [astro-ph]}}.

\bibitem{Binetruy:1996xj}
P.~Binetruy and G.~R. Dvali, ``{D term inflation},''
  \href{http://dx.doi.org/10.1016/S0370-2693(96)01083-0}{{\em Phys. Lett.}
  {\bfseries B388} (1996) 241--246},
\href{http://arxiv.org/abs/hep-ph/9606342}{{\ttfamily arXiv:hep-ph/9606342
  [hep-ph]}}.

\bibitem{Akrami:2018odb}
{\bfseries Planck} Collaboration, Y.~Akrami {\em et~al.}, ``{Planck 2018
  results. X. Constraints on inflation},''
\href{http://arxiv.org/abs/1807.06211}{{\ttfamily arXiv:1807.06211
  [astro-ph.CO]}}.

\bibitem{Stewart:1994pt}
E.~D. Stewart, ``{Mutated hybrid inflation},''
  \href{http://dx.doi.org/10.1016/0370-2693(94)01646-T}{{\em Phys. Lett.}
  {\bfseries B345} (1995) 414--415},
\href{http://arxiv.org/abs/astro-ph/9407040}{{\ttfamily arXiv:astro-ph/9407040
  [astro-ph]}}.

\bibitem{Lazarides:1995vr}
G.~Lazarides and C.~Panagiotakopoulos, ``{Smooth hybrid inflation},''
  \href{http://dx.doi.org/10.1103/PhysRevD.52.R559}{{\em Phys. Rev.} {\bfseries
  D52} (1995) R559--R563},
\href{http://arxiv.org/abs/hep-ph/9506325}{{\ttfamily arXiv:hep-ph/9506325
  [hep-ph]}}.

\bibitem{Reece:2018zvv}
M.~Reece, ``{Photon Masses in the Landscape and the Swampland},''
  \href{http://dx.doi.org/10.1007/JHEP07(2019)181}{{\em JHEP} {\bfseries 07}
  (2019) 181},
\href{http://arxiv.org/abs/1808.09966}{{\ttfamily arXiv:1808.09966 [hep-th]}}.

\bibitem{Witten:1978bc}
E.~Witten, ``{Instantons, the Quark Model, and the 1/n Expansion},''
\href{http://dx.doi.org/10.1016/0550-3213(79)90243-8}{{\em Nucl. Phys.}
  {\bfseries B149} (1979) 285}.

\bibitem{Witten:1980sp}
E.~Witten, ``{Large N Chiral Dynamics},''
\href{http://dx.doi.org/10.1016/0003-4916(80)90325-5}{{\em Ann. Phys.}
  {\bfseries 128} (1980) 363}.

\bibitem{Witten:1998uka}
E.~Witten, ``{Theta dependence in the large N limit of four-dimensional gauge
  theories},'' \href{http://dx.doi.org/10.1103/PhysRevLett.81.2862}{{\em Phys.
  Rev. Lett.} {\bfseries 81} (1998) 2862--2865},
\href{http://arxiv.org/abs/hep-th/9807109}{{\ttfamily arXiv:hep-th/9807109}}.

\bibitem{Lyth:1999sp}
D.~H. Lyth, ``{The Parameter space for tree level hybrid inflation},''
\href{http://arxiv.org/abs/hep-ph/9904371}{{\ttfamily arXiv:hep-ph/9904371
  [hep-ph]}}.

\bibitem{Coleman:1973jx}
S.~R. Coleman and E.~J. Weinberg, ``{Radiative Corrections as the Origin of
  Spontaneous Symmetry Breaking},''
\href{http://dx.doi.org/10.1103/PhysRevD.7.1888}{{\em Phys. Rev.} {\bfseries
  D7} (1973) 1888--1910}.

\bibitem{Kaloper:2015jcz}
N.~Kaloper, M.~Kleban, A.~Lawrence, and M.~S. Sloth, ``{Large Field Inflation
  and Gravitational Entropy},''
  \href{http://dx.doi.org/10.1103/PhysRevD.93.043510}{{\em Phys. Rev.}
  {\bfseries D93} no.~4, (2016) 043510},
\href{http://arxiv.org/abs/1511.05119}{{\ttfamily arXiv:1511.05119 [hep-th]}}.

\bibitem{Manohar:1983md}
A.~Manohar and H.~Georgi, ``{Chiral Quarks and the Nonrelativistic Quark
  Model},''
\href{http://dx.doi.org/10.1016/0550-3213(84)90231-1}{{\em Nucl. Phys.}
  {\bfseries B234} (1984) 189--212}.

\bibitem{Gavela:2016bzc}
B.~M. Gavela, E.~E. Jenkins, A.~V. Manohar, and L.~Merlo, ``{Analysis of
  General Power Counting Rules in Effective Field Theory},''
\href{http://arxiv.org/abs/1601.07551}{{\ttfamily arXiv:1601.07551 [hep-ph]}}.

\bibitem{Efstathiou:2005tq}
G.~Efstathiou and K.~J. Mack, ``{The Lyth Bound Revisited},''
  \href{http://dx.doi.org/10.1088/1475-7516/2005/05/008}{{\em JCAP} {\bfseries
  0505} (2005) 008},
\href{http://arxiv.org/abs/astro-ph/0503360}{{\ttfamily
  arXiv:astro-ph/0503360}}.

\bibitem{Kallosh:1995hi}
R.~Kallosh, A.~D. Linde, D.~A. Linde, and L.~Susskind, ``{Gravity and global
  symmetries},'' \href{http://dx.doi.org/10.1103/PhysRevD.52.912}{{\em Phys.
  Rev.} {\bfseries D52} (1995) 912--935},
\href{http://arxiv.org/abs/hep-th/9502069}{{\ttfamily arXiv:hep-th/9502069}}.

\bibitem{Kamionkowski:1992mf}
M.~Kamionkowski and J.~March-Russell, ``{Planck scale physics and the
  Peccei-Quinn mechanism},''
  \href{http://dx.doi.org/10.1016/0370-2693(92)90492-M}{{\em Phys. Lett.}
  {\bfseries B282} (1992) 137--141},
\href{http://arxiv.org/abs/hep-th/9202003}{{\ttfamily arXiv:hep-th/9202003}}.

\bibitem{Hebecker:2016dsw}
A.~Hebecker, P.~Mangat, S.~Theisen, and L.~T. Witkowski, ``{Can Gravitational
  Instantons Really Constrain Axion Inflation?},''
  \href{http://dx.doi.org/10.1007/JHEP02(2017)097}{{\em JHEP} {\bfseries 02}
  (2017) 097},
\href{http://arxiv.org/abs/1607.06814}{{\ttfamily arXiv:1607.06814 [hep-th]}}.

\bibitem{Daus:2020vtf}
T.~Daus, A.~Hebecker, S.~Leonhardt, and J.~March-Russell, ``{Towards a
  Swampland Global Symmetry Conjecture using Weak Gravity},''
\href{http://arxiv.org/abs/2002.02456}{{\ttfamily arXiv:2002.02456 [hep-th]}}.

\bibitem{Dimopoulos:2003iy}
S.~Dimopoulos and S.~D. Thomas, ``{Discretuum versus continuum dark energy},''
  \href{http://dx.doi.org/10.1016/j.physletb.2003.08.061}{{\em Phys. Lett.}
  {\bfseries B573} (2003) 13--19},
\href{http://arxiv.org/abs/hep-th/0307004}{{\ttfamily arXiv:hep-th/0307004
  [hep-th]}}.

\bibitem{Julia:1979ur}
B.~Julia and G.~Toulouse, ``{The Many Defect Problem: Gauge Like Variables for
  Ordered Media Containing Defects},''
\href{http://dx.doi.org/10.1051/jphyslet:019790040016039500}{{\em J. Phys.
  Lett.} {\bfseries 40} (1979) 396}.

\bibitem{Aurilia:1980xj}
A.~Aurilia, H.~Nicolai, and P.~K. Townsend, ``{Hidden Constants: The Theta
  Parameter of QCD and the Cosmological Constant of N=8 Supergravity},''
\href{http://dx.doi.org/10.1016/0550-3213(80)90466-6}{{\em Nucl. Phys.}
  {\bfseries B176} (1980) 509--522}.

\bibitem{Dvali:2003br}
G.~Dvali and A.~Vilenkin, ``{Cosmic attractors and gauge hierarchy},''
  \href{http://dx.doi.org/10.1103/PhysRevD.70.063501}{{\em Phys. Rev.}
  {\bfseries D70} (2004) 063501},
\href{http://arxiv.org/abs/hep-th/0304043}{{\ttfamily arXiv:hep-th/0304043
  [hep-th]}}.

\bibitem{Dvali:2005an}
G.~Dvali, ``{Three-form gauging of axion symmetries and gravity},''
\href{http://arxiv.org/abs/hep-th/0507215}{{\ttfamily arXiv:hep-th/0507215}}.

\bibitem{Brown:1987dd}
J.~D. Brown and C.~Teitelboim, ``{Dynamical neutralization of the cosmological
  constant},''
\href{http://dx.doi.org/10.1016/0370-2693(87)91190-7}{{\em Phys. Lett.}
  {\bfseries B195} (1987) 177--182}.

\bibitem{Brown:1988kg}
J.~D. Brown and C.~Teitelboim, ``{Neutralization of the Cosmological Constant
  by Membrane Creation},''
\href{http://dx.doi.org/10.1016/0550-3213(88)90559-7}{{\em Nucl. Phys.}
  {\bfseries B297} (1988) 787--836}.

\bibitem{Bousso:2000xa}
R.~Bousso and J.~Polchinski, ``{Quantization of four-form fluxes and dynamical
  neutralization of the cosmological constant},'' {\em JHEP} {\bfseries 06}
  (2000) 006,
\href{http://arxiv.org/abs/hep-th/0004134}{{\ttfamily arXiv:hep-th/0004134}}.

\bibitem{Marchesano:2014mla}
F.~Marchesano, G.~Shiu, and A.~M. Uranga, ``{F-term Axion Monodromy
  Inflation},''
\href{http://arxiv.org/abs/1404.3040}{{\ttfamily arXiv:1404.3040 [hep-th]}}.

\bibitem{Hebecker:2014eua}
A.~Hebecker, S.~C. Kraus, and L.~T. Witkowski, ``{D7-Brane Chaotic
  Inflation},''
\href{http://arxiv.org/abs/1404.3711}{{\ttfamily arXiv:1404.3711 [hep-th]}}.

\bibitem{Blumenhagen:2014gta}
R.~Blumenhagen and E.~Plauschinn, ``{Towards Universal Axion Inflation and
  Reheating in String Theory},''
\href{http://arxiv.org/abs/1404.3542}{{\ttfamily arXiv:1404.3542 [hep-th]}}.

\bibitem{Grimm:2014vva}
T.~W. Grimm, ``{Axion Inflation in F-theory},''
\href{http://arxiv.org/abs/1404.4268}{{\ttfamily arXiv:1404.4268 [hep-th]}}.

\bibitem{Dudas:2014pva}
E.~Dudas, ``{Three-form multiplet and Inflation},''
  \href{http://dx.doi.org/10.1007/JHEP12(2014)014}{{\em JHEP} {\bfseries 12}
  (2014) 014},
\href{http://arxiv.org/abs/1407.5688}{{\ttfamily arXiv:1407.5688 [hep-th]}}.

\bibitem{Ibanez:2014swa}
L.~E. Ibanez, F.~Marchesano, and I.~Valenzuela, ``{Higgs-otic Inflation and
  String Theory},'' \href{http://dx.doi.org/10.1007/JHEP01(2015)128}{{\em JHEP}
  {\bfseries 01} (2015) 128},
\href{http://arxiv.org/abs/1411.5380}{{\ttfamily arXiv:1411.5380 [hep-th]}}.

\end{thebibliography}\endgroup

\end{document}